\documentclass[aps,prx,reprint,superscriptaddress]{revtex4-1}
\usepackage{amsmath}
\usepackage{amssymb}
\usepackage{CJK}
\usepackage{graphicx}
\usepackage{dcolumn}
\usepackage{bm}
\usepackage{graphicx} 
\usepackage{hyperref}
\usepackage[dvipsnames]{xcolor}
\usepackage[tight]{subfigure} 
\usepackage{mathtools}
\usepackage{bbold}
\usepackage{ulem}
\usepackage{tikz}

\newcommand{\x}{\bm{x}}
\newcommand{\y}{\bm{y}}
\renewcommand{\r}{\bm{r}}
\renewcommand{\k}{\bm{k}}
\newcommand{\R}{\bm{R}}
\newcommand{\q}{\bm{q}}
\newcommand{\Q}{\bm{Q}}
\newcommand{\K}{\bm{K}}
\renewcommand{\u}{\bm{u}}
\newcommand{\SU}{\rm{SU}}

\newcommand{\Rot}{R_{\frac{\pi}{2}}}
\newcommand{\RInv}{\tilde{R}_{\pi}}
\newcommand{\Tt}{\tilde{\mathcal{T}}}

\begin{document}
\begin{CJK*}{UTF8}{}

\title{Magic continuum in twisted bilayer square lattice with staggered flux}

\CJKfamily{gbsn}

\author{Zhu-Xi Luo (罗竹悉)}
\affiliation{Kavli Institute for Theoretical Physics, University of California, Santa Barbara, CA 93106, USA}

\author{Cenke Xu}
\affiliation{Department of Physics, University of California, Santa Barbara, CA 93106, USA}

\author{Chao-Ming Jian}
\affiliation{Department of Physics, Cornell University, Ithaca, New York 14853, USA}

\date{\today}

\begin{abstract}
We derive the general continuum model for a bilayer system of staggered-flux square lattices, with arbitrary elastic deformation in each layer. Applying this general continuum model to the case where the two layers are rigidly rotated relative to each other by a small angle, we obtain the band structure of the twisted bilayer staggered-flux square lattice. We show that this band structure exhibits a ``magic continuum" in the sense that an exponential reduction of the Dirac velocity and bandwidths occurs in a large parameter regime. We show that the continuum model of the twisted bilayer system effectively describes a massless Dirac fermion in a spatially modulating magnetic field, whose renormalized Dirac velocity can be exactly calculated. We further give an intuitive argument for the emergence of flattened bands near half filling  in the magic continuum and provide an estimation of the large number of associated nearly-zero-energy states. We also show that the entire band structure of the twisted bilayer system is free of band gaps due to symmetry constraints. 
\end{abstract}


\maketitle
\end{CJK*}

\section{Introduction}

The experimental observations of correlated insulating and superconducting behaviors in twisted bilayer graphene \cite{cao2018correlated,cao2018unconventional,yankowitz2019tuning,sharpe2019emergent,lu2019superconductors,serlin2020intrinsic,kerelsky2019maximized,xie2019spectroscopic,jiang2019charge} have generated the era of ``twistronics" \cite{carr2017twistronics}. Since then, moir\'e physics of various assembled atomically-thin systems has been explored, including transition metal dichalgonides heterostructures \cite{tran2019evidence,jin2019observation,seyler2019signatures,alexeev2019resonantly,tang2020simulation,regan2020mott,shimazaki2020strongly,wang2020correlated}, other graphene-based heterostructures such as twisted double bilayer graphene \cite{liu2020tunable,burg2019correlated,cao2020tunable,shen2020correlated}, ABC-stacked trilayer graphene/boron nitride \cite{chen2019evidence,chen2019signatures,chen2020tunable}, twisted monolayer-bilayer graphene \cite{chen2020electrically}, and twisted trilayer graphene \cite{tsai2019correlated}. There have also been theoretical proposals of other exotic moir\'e systems, for example,
bilayers of general Bravais lattices \cite{kariyado2019flat}, van der Waals magnets \cite{hejazi2020noncollinear}, superconductors \cite{can2020high,volkov2020magic}, gapped spin liquid \cite{may2020twisted}, surface states of topological insulators \cite{cano2020moir,wang2020moir}, and cold atomic systems \cite{gonzalez2019cold,fu2020magic, salamon2020simulating,luo2021spin}, to name a few.

The correlated behaviors in 
twisted bilayer graphene are known to be associated with the flattening of bands near the charge neutrality. Such band flattening was theoretically predicted \cite{bistritzer2011moire,bistritzer2011moire1,dos2012continuum,shallcross2010electronic} and experimentally observed \cite{yin2015experimental,cao2018correlated,cao2018unconventional}. In fact, the widths of the energy bands near the charge neutrality are highly sensitive to the twist angles in the twisted bilayer graphene system. The flat bands only occurs inside very narrow windows around certain discrete values of magic angles \cite{bistritzer2011moire}. However, it is experimentally challenging to precisely control the twist angles between the two graphene sheets. Different samples tend to settle into configurations with different twist angles or even spatially inhomogeneous twist angles. The high sensitivity of the electronic structure makes it difficult to interpret experimental measurements of correlated physics and understand their underlying mechanisms in such systems.
It would then be much more ideal if the band flattening happens in a wide range of twist angles. An interesting example of such scenario is given by twisted bilayer WSe$_2$ which is experimentally shown to exhibit band flattening and associated correlated physics in over a continuum range of twist angles. This continuum range of twist angles is referred to as the magic continuum \cite{wang2020correlated}. However, we note that gapless Dirac cones, a prominent feature of the electronic structure of twisted bilayer graphene, is not present in the twisted bilayer WSe$_2$ system. It is interesting to search for systems with both gapless Dirac cones near charge neutrality and a magic continuum of twist angles where band flattening occurs.

The gapless Dirac cones of the twisted bilayer graphene are inherited from the those of each individual graphene sheet. As a generalization, it is natural to consider twisted bilayer systems where Dirac cones are present in the electronic structure of each individual layer. In this work, we will focus on the twisted bilayer system consisting of two layers of staggered-flux square lattice. A single-layer staggered-flux square lattice describes a tight-banding model with nearest-neighbor hoppings on the square lattice subject to a staggered magnetic flux pattern. The band structure of this tight-binding model contains two gapless Dirac cones at half filling. This staggered-flux square-lattice model was initially proposed to capture the band structure of fractionalized particles in underdoped cuprates \cite{anderson1987resonating,affleck1988large,marston1989large,wen1996theory,kim1999theory} (see [\onlinecite{lee2006doping}] for a review). It has also been widely investigated as a prominent mean-field ansatz for an algebraic quantum spin liquid \cite{rantner2001electron,wen2002quantum,rantner2002spin,hermele2004stability,Hermele} that can be viewed as a parent state of many competing orders \cite{Hermele}. In the contexts of both cuprates and spin liquids, it is the fractionalized particles (i.e. spinons) that experience the staggered flux on the square lattice. In our work, we will focus on the case where the staggered-flux square-lattice tight-binding model describes the hopping of electrons within each layer of our bilayer system. 
The examination of the moir\'e physics of bilayer staggered-flux square lattice will also serve as a preceding step towards understanding the physics of twisted bilayer of algebraic spin liquids.

In the seminal work \cite{bistritzer2011moire}, a continuum model was developed to describe the band structure of twisted bilayer graphene. This continuum model and its generalizations have been the foundation of the theoretical studies of twisted bilayer graphene and other moir\'e systems. The method based on the continuum models has proven to be advantageous for general moir\'e systems: it restores periodicity in a quasiperiodic system, reduces the number of dimensionless parameters, and is flexible to incorporate general smooth deformations. Following the method introduced in Ref. \onlinecite{bistritzer2011moire}, the continuum models of bilayer or multi-layer moir\'e systems can be obtained by studying reasonable forms of interlayer tunnelings in the momentum space. Recently, a new real-space derivation of the continuum model for twisted bilayer graphene has been developed \cite{Balents}. In this new derivation, the original continuum model of twisted bilayer graphene was directly obtained from symmetry-based bootstrap analysis. More generally, as shown in Ref. \onlinecite{Balents}, the same method enables the derivation of the continuum model of general bilayer graphene systems with arbitrary independent elastic deformation in each graphene layer. In this work, we follow the real-space symmetry-based method to construct the continuum model for arbitrary elastically deformed bilayer of the staggered-flux square lattice. 

A case of particular interest is the twisted bilayer staggered-flux square lattice system where the two layers of square lattices are rigidly rotated relative to each other by a small twist angle. We solve the corresponding continuum model for the band structure and find that gapless Dirac cone exists near half filling with the Dirac velocity is substantially renormalized compared to that in a single-layer staggered-flux square lattice. In particular, it decreases exponentially as the interlayer tunneling increases and/or as the the twist angle decreases. Indicated by the drastic reduction of the Dirac velocity, the flattening of the bands near the half filling occurs in a large regime of the tunneling parameter and twist angles, i.e., there is a magic continuum. This magic continuum can be understood as arising from Dirac fermions subject to a spatially periodically modulated effective magnetic field induced by the interlayer tunneling. Also, we show that band structure of the twisted bilayer staggered-flux square lattice system  is free of band gaps at any energy. The symmetries of system enforce all the bands to be connected to the neighboring ones. This infinite connectivity of all the bands in the twisted bilayer staggered-flux square lattice system is similar to the ``perfect metal" discussed in graphene-based heterostructures \cite{mora2019flatbands,Song2020TBGII}. 

The remainder of this paper is organized as follows. In Sec. \ref{sec:single}, we review the tight-biding model of a single-layer staggered-flux square lattice and its low-energy continuum model for the band structure. We further extend the continuum model to incorporate an arbitrary elastic deformation of the square lattice. In Sec. \ref{sec:bilayer}, we consider a bilayer systems with two elastically deformed staggered-flux square lattice layers. We present the symmetry-based bootstrap analysis of the general form of the Hamiltonian of the bilayer system in Sec. \ref{subsec:bootstrap}. In Sec. \ref{subsec:uniform} and Sec. \ref{subsec:twist}, we apply the general Hamiltonian to the two special cases respectively: (1) the bilayer systems with the two square-lattice layers rigidly shifted relative to each other and (2) the twisted bilayer system with the two square-lattice layers rigidly rotated relative to each other by a small twist angle. We also present the numerical calculation of band structure of the twisted bilayer system. In Sec. \ref{sec:magic}, we perform more detailed analytical study on the band structure of twisted bilayer system. In particular, we show analytically the drastic reduction of Dirac velocity as we increase the interlayer tunneling (or decrease the twist angle). We present an intuitive argument for the emergence of flattened bands and the large number of associated low-energy states in the same parameter regime. We also discuss a symmetry-based argument that enforces the infinite connectivity of entire band structure.
We then conclude with some extensions and outlook in Sec. \ref{sec:discuss}. 

\section{Single-layer continuum model}
\label{sec:single}

In this section, we first review the basics of the staggered-flux square-lattice tight-binding model, and then derive a continuum model that incorporates an arbitrary smooth lattice deformation. 

\subsection{Review of staggered-flux square-lattice model}
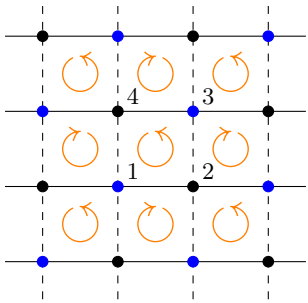
\begin{figure}[htbp]
\centering
\begin{tikzpicture}
\foreach \Y in {-1,0,...,2}
{\draw (-1.5,\Y)--(2.5,\Y);}
\foreach \X in {-1,0,...,2}
{\draw[dashed] (\X,-1.5)--(\X,2.5);}
\filldraw[blue] (0,0) circle (2pt);
\node at (0.2,0.2) {$1$};
\filldraw (1,0) circle (2pt);
\node at (1.2,0.2) {$2$};
\filldraw (0,1) circle (2pt);
\node at (0.2,1.2) {$4$};
\filldraw[blue] (1,1) circle (2pt);
\node at (1.2,1.2) {$3$};
\filldraw[blue] (-1,-1) circle (2pt);
\filldraw[blue] (1,-1) circle (2pt);
\filldraw[blue] (2,0) circle (2pt);
\filldraw[blue] (0,2) circle (2pt);
\filldraw[blue] (2,2) circle (2pt);
\filldraw[blue] (-1,1) circle (2pt);
\filldraw (-1,0) circle (2pt);
\filldraw (-1,2) circle (2pt);
\filldraw (0,-1) circle (2pt);
\filldraw (1,2) circle (2pt);
\filldraw (2,-1) circle (2pt);
\filldraw (2,1) circle (2pt);
\node at (0.5,0.5) {\huge \textcolor{orange}{$\circlearrowleft$}};
\node at (1.5,0.5) {\huge \textcolor{orange}{$\circlearrowright$}};
\node at (-0.5,0.5) {\huge \textcolor{orange}{$\circlearrowright$}};
\node at (0.5,1.5) {\huge \textcolor{orange}{$\circlearrowright$}};
\node at (1.5,1.5) {\huge \textcolor{orange}{$\circlearrowleft$}};
\node at (-0.5,1.5) {\huge \textcolor{orange}{$\circlearrowleft$}};
\node at (0.5,-0.5) {\huge \textcolor{orange}{$\circlearrowright$}};
\node at (1.5,-0.5) {\huge \textcolor{orange}{$\circlearrowleft$}};
\node at (-0.5,-0.5) {\huge \textcolor{orange}{$\circlearrowleft$}};
\end{tikzpicture}
\caption{The square lattice with staggered flux is shown. The blue/black dots correspond to the $A$/$B$ sublattice respectively. The hopping amplitudes from sublattices $A$ to $B$ through solid black links are given by $(it+\Delta)$, while those through the dashed links are given by $(it-\Delta)$. The neighboring plaquettes thus host staggered flux $\pm\Phi$. To obtain the spectrum, we consider a 4-site unit cell with the four sites within a unit cell labeled as shown above.}
\label{fig:uc}
\end{figure}

The Hamiltonian of the staggered-flux square-lattice model describes spinful fermions hopping on the square lattice that is divided into two sublattices $A$ and $B$: 
\begin{equation}
H=-\sum_{\r\in A} \sum_{\r'\in n.n.} [ (i t+(-1)^{r_y-r_y'}\Delta)f_{\r \alpha}^\dagger f_{\r'\alpha}+h.c.].
\label{eq:lattice_ham}
\end{equation}
where $\sum_{\r\in A}$ is a summation over the sites in the sublattice $A$ and $\sum_{\r'\in n.n.}$ sums over the $B$-sublattice sites $\r'$ that are the nearest neighbors of $\r$. The subscript $\alpha$ is the spin index. The hopping amplitudes are shown in Fig. \ref{fig:uc}.
This Hamiltonian describes fermions hopping in a background of staggered magnetic flux. The flux though each square plaquette is given by $\pm\Phi=\pm 4\arctan (t/\Delta)$. The signs of fluxes are opposite for neighboring plaquettes.

The naive translations by one site along the $x$- and $y$-direction do not leave the Hamiltonian in Eq. \eqref{eq:lattice_ham} invariant. Such translations, when combined with an extra particle-hole transformation, become the symmetries of the Hamiltonian. We denote these spatial translations followed by a particle-hole transformation as $T_x$ and $T_y$. Similarly, the Hamiltonian has a mirror symmetry $\mathcal{M}_x$ that maps the site at $\r = (r_x, r_y )$ to the site at $\mathcal{M}_x \r \equiv (-r_x, r_y)$ and, at the same time, maps particles to holes. Moreover, the Hamiltonian has a time-reversal symmetry $\mathcal{T}$ and a plaquette-centered four-fold spatial rotation symmetry $\Rot$ that transform the sites following $\r \rightarrow \Rot \r \equiv (-r_y+1, r_x)$. The actions of these symmetries on the lattice fermions are given by 
\begin{equation}
\begin{split}
& T_x:\quad  f_{\r, \alpha}\rightarrow \epsilon_{\r}(i\sigma^2)_{\alpha\beta} f_{\r+\hat{\x},\beta}^\dagger,
\\
& T_y:\quad  f_{\r, \alpha}\rightarrow  \epsilon_{\r}(i\sigma^2)_{\alpha\beta} f_{\r+\hat{\y},\beta}^\dagger,
\\
& \mathcal{M}_x:\quad   f_{\r,\alpha}\rightarrow f_{\mathcal{M}_x\r,\alpha},
\\
& \Rot: \quad f_{\r,\alpha}\rightarrow \epsilon_{\r}f_{\Rot\r,\alpha},
\\
& \mathcal{T}:\quad f_{\r,\alpha}\rightarrow \epsilon_{\r} f_{\r,\alpha}^\dagger,
\end{split}
\label{eq:micro_sym}
\end{equation}
where $\bm{\sigma} = (\sigma^1, \sigma^2, \sigma^3)$ acts in the $\SU(2)$-spin space and $\epsilon_{\r}$ takes value $1$ for sublattice $A$ and $- 1$ for sublattice $B$. The transformation laws for $f^\dagger$'s can be obtained by taking hermitian conjugate on both sides of the transformation laws above in \eqref{eq:micro_sym}.
In the context of algebraic spin liquids, these transformations were initially introduced as the projective symmetry group acting on the fermionic spinons that couples to dynamical gauge fields \cite{wen2002quantum}. In this work, we view these transformations as the actual symmetry action on gauge-neutral fermions hopping on the staggered-flux square lattice.

To obtain the band structure of the Hamiltonian \eqref{eq:lattice_ham},  we take a four-site unit cell on the square lattice as shown in Fig. \ref{fig:uc}, following the convention in Ref. \onlinecite{Hermele}. Each unit cell is assigned a coordinate $\R = (R_x, R_y)$ with both $R_x$ and $R_y$ even integers. The 4 sites labeled by the $i=1,2,3,4$ are located at $\r(\R,i)= \R + \bm{v}_i$ where
\begin{equation}
\bm{v}_i=\begin{cases}
0, & i=1\\
\hat{\x}, & i=2\\
\hat{\x}+\hat{\bm{y}}, & i=3\\
\hat{\bm{y}}, & i=4.
\end{cases}
\label{eq:sublattice}
\end{equation}
The sublattice $A$ corresponds to $i=1,3$ while the sublattice $B$ corresponds to $i=2,4$. The fermion operator at $\r(\R,i)$ will be denoted as $f_{\R i\alpha}$.

\begin{figure}[htbp]
\centering
\includegraphics[scale=0.5]{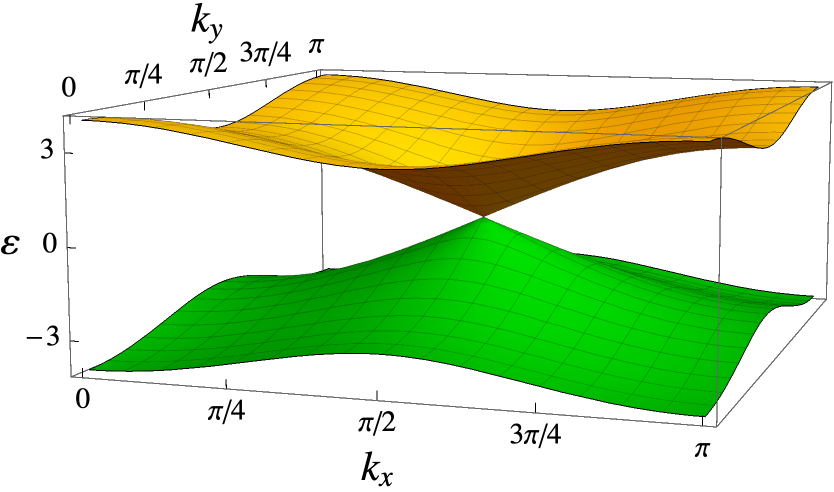}
\caption{Spectrum of the monolayer staggered-flux square lattice with $t=\Delta=1$. There are two degenerate Dirac cones for each spin located at $\K=(\pi/2,\pi/2)$ in the reduced Brillouin zone.}
\label{fig_mono_dispersion}
\end{figure}

The energy spectrum of the model Eq. \eqref{eq:lattice_ham} is given by 
\begin{equation}
\begin{split}
\varepsilon(\k)=\pm 2 \Big[ & \pm 2(\Delta^2-t^2)\cos k_x\cos k_y \\
&  +\frac{1}{2}(\Delta^2+t^2)(2+\cos 2k_x+\cos 2k_y)\Big]^{1/2},
\end{split}
\label{eq:mono_dispersion}
\end{equation}
where the two $\pm$ signs are independent of each other. The four combination of $\pm$ signs corresponds to four different bands. In the reduced Brillouin zone $k_x, k_y\in [0,\pi)$, there is a gapless point at $\K=(\pi/2,\pi/2)$ at half filling, as shown in Fig. \ref{fig_mono_dispersion}. In this model, half filling occurs at zero energy due to the symmetry $\mathcal{T}$. One observes that the position of the gapless point is independent of $t/\Delta$. This gapless point can be captured by two gapless Dirac cones in a continuum description. The ratio $t/\Delta$ controls the Dirac velocity anisotropy of these Dirac cones. In this work, we will focus on the isotropic limit where $t/\Delta = 1$, but the main features of our final results remain robust when anisotropy is present \footnote{As a side note, in the context of the algebraic spin liquid, the Dirac velocity anisotropy for the spinons, which is controlled by the deviation of $t/\Delta$ from 1, has been shown to be irrelevant in the renormalization group sense in the large-$N_f$ limit \cite{vafek2002relativity, franz2002qed, Hermele}. Here, $N_f$ refers to the number of flavor of spinons.}. 

To obtain the continuum model for the gapless Dirac cones at half-filling, we expand the Hamiltonian around the $\K$ point and introduce the Dirac fermions basis $\psi_{a\alpha}^I (\R)$ in the following way: 
\begin{equation}
 \resizebox{.95\hsize}{!}{$\left(\begin{matrix} f_{\R 1\alpha} \\ f_{\R 2\alpha} \\ f_{\R 3\alpha} \\ f_{\R 4\alpha} \end{matrix}\right) \\ 
\sim 
 e^{-i\K\cdot\R}~ \left(\begin{matrix} 1 & 0 & 0 & e^{-3i\pi/4}\\ 0 & i & e^{-3i\pi/4} & 0\\ 1 & 0 & 0 & e^{i\pi/4} \\ 0 & -i & e^{-3i\pi/4} & 0  \end{matrix}\right)
\left(\begin{matrix} \psi_{1\alpha}^1 \\ \psi_{1\alpha}^2 \\ \psi_{2\alpha}^1 \\ \psi_{ 2\alpha}^2 \end{matrix}\right)(\R).$}
\label{eq:f_psi}
\end{equation}
Here the superscript $I=1,2$ of $\psi^{I}_{a\alpha}$ labels the two components of a Dirac spinor, and the subscript $a=1,2$ and $\alpha=1,2$ label the valley and spin degrees of freedom respectively. Denoting the deviation of momentum $\k$ from the $\K$ point as $\q$, i.e. $\q\equiv \k-\K$, and further switching to the $45^\circ$-rotated coordinates,
$q_1=\frac{1}{\sqrt{2}}(q_x+q_y),$ $q_2=\frac{1}{\sqrt{2}} (-q_x+q_y),$
the continuum Hamiltonian becomes
\begin{equation}
H=\int \frac{d^2\q}{(2\pi)^2}~\psi^\dagger(\q)(q_1\tau^1+q_2\tau^2)\psi(\q),
\label{eq:Ham}
\end{equation}
where we have set the Dirac velocity to be $1$ and $\bm{\tau}=(\tau^1, \tau^2, \tau^3)$ are the Pauli matrices acting in the Dirac spinor space, the space indexed by the superscript $I=1,2$ of the Dirac fermions $\psi^I_{a\alpha}$. We also introduce the Pauli matrices $\bm{\mu} = (\mu^1, \mu^2, \mu^3)$ that act on the two-fold valley space. The Pauli matrices $\bm{\mu}$ generate the SU$(2)_{\rm valley}$ rotation of the Dirac fermions $\psi$. The continuum Hamiltonian manifestly has the $\SU(4)\supset \SU(2)_{\text{spin}}\times \SU(2)_{\text{valley}}$ symmetry generated by $\{\sigma^i, \mu^i, \sigma^i\mu^j\}$. Here, remember that $\bm{\sigma}$ generates the SU$(2)_{\rm spin}$ rotation of the Dirac fermions. 

In the following, we will suppress the Dirac spinor index $I$, valley index $a$ and the spin index $\alpha$ and write $\psi$ as the shorthand notation for the 8-component Dirac spinor $\psi^I_{a\alpha}$. The symmetries listed in Eq.  \eqref{eq:micro_sym} then act on the Dirac fermions as
\begin{equation}
\begin{split}
& T_x:\quad \psi(\R) \rightarrow \sigma^2 \tau^1 \psi^*(\R), \\
& T_y:\quad \psi(\R)\rightarrow -\sigma^2\mu^3\tau^1\psi^*(\R),  \\
& \mathcal{M}_x:\quad \psi(\R)\rightarrow W_1 \psi(\mathcal{M}_x\R)\\
& \Rot:\quad  \psi(\R)\rightarrow W_2\psi(\Rot\R),\\
& \mathcal{T}:\quad \psi(\R)\rightarrow -\mu^3\tau^3\psi^*(\R),
\end{split}
\label{eq:sym_psi_R}
\end{equation}
where $W_1=(-i\mu^2)\exp\left[\frac{i\pi}{2}\left(\frac{\tau^1+\tau^2}{\sqrt{2}}\right)\right]$ and $W_2=e^{3i\pi/4}\exp\left[\frac{i\pi}{2}\left(\frac{\mu^1+\mu^2}{\sqrt{2}}\right)\right]\exp\left(\frac{i\pi}{4}\tau^3\right)$. The transformation laws for $\psi^\dagger$'s can be obtained by taking the Hermitian conjugate on both sides of the transformation laws in Eq. \eqref{eq:sym_psi_R}.

\subsection{Elastic deformation}
Now we move on to the elastically deformed version of the staggered-flux square lattice. The derivation of the continuum Hamiltonian closely follows the formalism developed in Ref. \onlinecite{Balents} in the context of twisted bilayer graphene. 

In the following, we will make use of the Eulerian coordinates
\begin{equation}
\x=\R+\u(\x),
\label{eq:Eulerian}
\end{equation}
where $\bm x$ is the spatial coordinate of sites in the deformed lattice, $\R$ is the coordinate of the same lattice site prior to the deformation and 
$\u$ describes the deformation. We will assume that $\partial \u \ll 1$, such that the elasticity theory applies. Notice that $\u(\x)$ and $\R$ are both treated as functions of the Eulerian coordinate $\x$. An alternative choice of coordinate system is the Lagrangian coordinate where $\x$ and $\u$ are both treated as functions of $\R$, the spatial coordinate of the pre-deformed lattice site. We choose to use the Eulerian coordinate $\x$ over the Lagrangian coordinate $\R$ for the purpose of our later discussion on the bilayer system with the two layers independently deformed. Two points in the two layers that share the same Lagrangian coordinate $\R$ can be far part due to the independent deformation in each layer. Therefore, the spatial locality in the deformed bilayer system is not manifest in the Lagrangian coordinate $\R$. In contrast, being the real-space coordinates of lattice sites after the deformation, the Eulerian coordinate $\x$ avoids this problem. 

The fermion modes $f_{\R i \alpha}$, its associated Hamiltonian and symmetry transformations introduced in Sec. \ref{sec:single} are all formulated in the Lagrangian coordinate $\R$. In the following, we will suppress the subscripts $i$ and $\alpha$ in the fermion operator $f_{\R}$ for simplicity. In the presence of elastic deformation, we should view Eq. \eqref{eq:Eulerian} as a coordinate transformation that induces a new set of fermion operators $f_{\x} = |\det (\partial R_\mu)/(\partial x_\nu)|^{1/2} f_{\R(\x)}  \sim (1-\bm{\nabla}\cdot\u)^{1/2} e^{-i\K\cdot(\x-\u(\x))} U \psi (\R(\x))$ , where the Jacobian $|\det (\partial R_\mu)/(\partial x_\nu)|^{1/2}$ is required to ensure the correct fermionic anti-commutation relations and $U$ is the $4\times 4$ matrix introduced in Eq. \eqref{eq:f_psi} which only acts on the valley and Dirac spinor indices (but not the spin index) of $\psi$. We can define the Dirac fermion operator $\psi(\x)$ in the Eulerian coordinate via 
$
f_{\x}\sim e^{-i\K\cdot\x } U \psi (\x).
$ leading to the relation: 
\begin{equation}
\psi(\R)=(1-\bm{\nabla}\cdot\u)^{-1/2}e^{-i\K\cdot\u}\psi(\x) .
\label{eq:Rr}
\end{equation}
Note that the Dirac fermion operator $\psi(\x)$ in the Eulerian coordinate shares the same Dirac spinor, valley and spin indices as its Lagrangian-coordinate counterpart $\psi(\R)$.
Furthermore, under the coordinate change Eq. \eqref{eq:Eulerian}, the integration measure and the derivative change as $d^2\R\approx d^2\x (1-\bm{\nabla}\cdot\u)$ and $\quad \partial/{\partial R_i}\approx {\partial}/{\partial x_i}+({\partial u_j}/{\partial x_i})({\partial}/{\partial x_j})$ up to the first order in the derivative of $\u$. Now, we can rewrite the continuum Hamiltonian Eq. \eqref{eq:Ham} currently defined on a deformed lattice as a continuum Hamiltonian in the real space (parameterized by the Eulerian coordinate $\x$):
\begin{equation}
H_0=-i \int d^2\x\ \psi^\dagger (\x) [\tau^i\partial_i+\tau^j(\partial_j u^i)\partial_i-i\tau^iK_j \partial_iu^j]\psi(\x).
\label{eq:H_x}
\end{equation}
Here, we have only kept the terms up to the first-order derivative of $\u$. In this equation, $\partial_i=\partial/\partial x_i$ with $i=1,2$ are the derivatives with respect to the $45^\circ$-rotated version of the coordinate $\x$. 
We remark that the second term $\psi^\dagger \tau^j(\partial_j u^i)\partial_i \psi$ captures the rotation of the Dirac cone under the deformation while the third term $K_i\bm{\nabla}u_i$ in this equation captures the shift of the Dirac point in the momentum space. 

Each symmetry listed in Eq. \eqref{eq:sym_psi_R} leads to a symmetry in the continuum theory Eq. \eqref{eq:H_x} of the deformed lattice. Due to the change of coordinates Eq. \eqref{eq:Eulerian}, the form of symmetry actions on the Dirac fermion $\psi(\x)$ will be different from Eq. \eqref{eq:sym_psi_R}. Moreover, for the space-group symmetries including the mirror symmetry $\mathcal{M}_x$ and the four-fold rotation symmetry $\Rot$, the continuum theory should only be invariant under the simultaneous transformation of both the Dirac fermion $\psi(x)$ and the deformation field $\u(\x)$. Even though the symmetries $T_{x,y}$ are also space-group symmetry at the lattice scale, they should be viewed as ``internal symmetries" that only act on the Dirac spinor and the valley indices in the continuum theory. Therefore, $T_{x,y}$ should not involve any non-trivial action on the deformation field $\u(\x)$. The time-reversal symmetry $\mathcal{T}$ should also keep the deformation field $\u(\x)$ invariant.

Now we derive the form of symmetry actions for the continuum model Eq. $\eqref{eq:H_x}$.
Take $T_x$ as an example: if we were to take the same transformation law as that in \eqref{eq:sym_psi_R}, then among the three terms in the square bracket of \eqref{eq:H_x}, the first two terms are invariant, while the third term changes by a sign. To compensate for this sign change, we introduce an additional phase to the transformation,
\begin{equation}
T_x:\quad \psi(\x)\rightarrow e^{2i\K\cdot \u(\x)}\sigma^2\tau^1\psi^*(\x).
\label{eq:Tx_x}
\end{equation}
This extra phase will give an additional contribution to the first term in \eqref{eq:H_x}, thereby keeping the full Hamiltonian invariant (without additional transformation on $\u(\x)$). Physically, this extra phase factor reflects the shift of the Dirac cones in the momentum space introduced by the deformation $\u(\x)$. In the presence of the deformation $\u(\x)$, all the symmetry transformations of the fermion operators are summarized as follows
\begin{equation}
\begin{split}
& T_x:\quad \psi(\x)\rightarrow e^{2i\K\cdot \u(\x)}\sigma^2\tau^1\psi^*(\x), \\
& T_y:\quad \psi(\x)\rightarrow e^{2i\K\cdot \u(\x)}(-\sigma^2)\mu^3\tau^1\psi^*(\x)\\
& \mathcal{M}_x:\quad \psi(\x)\rightarrow e^{i (\K'-\K) \cdot \u ( \mathcal{M}_x \x)}W_1\psi(\mathcal{M}_x \x)\\
& \Rot:\quad \psi(\x)\rightarrow e^{i(\K''-\K)\cdot \u(\Rot \x)} W_2\psi(\Rot \x)\\
& \mathcal{T}:\quad \psi(\x)\rightarrow -\mu^3\tau^3\psi^*(\x),\quad i\rightarrow -i.
\end{split}
\label{eq:sym_x}
\end{equation}
where $\K' \equiv \mathcal{M}_x \K$ and $\K'' \equiv \Rot \K$. It turns out that $\K' = \K''$. For the symmetries $\mathcal{M}_x$ and $\Rot$, the deformation field must also undergoes the transformation $\u(\x)\rightarrow \hat{O}^{-1} \u(\hat{O}\x)$. Here, the $\mathcal{M}_x$ and $\Rot$ actions on a vector ${\bm v}$ are given by $\mathcal{M}_x: {\bm v} = (v_x,v_y) \rightarrow \mathcal{M}_x{\bm v} = (-v_x, v_y)$ and $\Rot: {\bm v} = (v_x,v_y) \rightarrow \Rot{\bm v} = (-v_y, v_x)$. The symmetry transformation law for $\psi^\dag$'s  can be obtained by taking the Hermitian conjugate on both sides of the transformation laws  in Eq. \eqref{eq:sym_x}

\section{Continuum model for bilayer staggered-flux square lattice with general deformations}
\label{sec:bilayer}

In this section, we consider two layers of staggered-flux square lattices with general deformations $\u_t(\x)$, $\u_b(\x)$ for the top and the bottom layers respectively. As is pointed out in Ref. \onlinecite{Balents}, one can start by considering the general form of interlayer tunneling:
\begin{equation}
H_1= \int d^2\x\ \psi_{b}^\dagger(\x) M[\u_t(\x),\u_b(\x)]\psi_{t}(\x)+h.c.,
\label{eq:H_1}
\end{equation}
Here, $\psi_{t,b}$ refers to the Dirac fermions in the two layers of the staggered-flux square lattices with their subscripts the layer index. Each of $\psi_{t}, \psi_b $ is an 8-component Dirac fermion with suppressed Dirac spinor, valley and spin indices. In this work, We assume spin-independent interlayer tunneling. Therefore, the interlayer tunneling matrix elements can be organized into a $4\times 4$ matrix $M$ that only act on the four-fold space labeled by the Dirac spinor index and the valley index. The form of Eq. \eqref{eq:H_1} guarantees the locality of the interlayer tunneling, which is an natural expectation for the continuum model. 
All the subdominant terms depending on the gradients of the displacements and/or the gradients of the Dirac fermion fields $\psi_{t,b}$ have been omitted. Now we use the symmetries given in Eq. \eqref{eq:Tx_x} and Eq. \eqref{eq:sym_x} to bootstrap the general form of $M$. 

\subsection{Bootstrap}
\label{subsec:bootstrap}

First, the deformation of both layers by a uniform vector should not change the interlayer physics. Therefore $M[\u_t,\u_b]=M[\u_t-\u_b]\equiv M[\u]$, where $\u \equiv \u_t - \u_b$. Second, deformation of a single layer by two lattice vectors should leave the physics invariant, which is due to the unit cell structure given by $\R= (R_x, R_y)$ with $R_{x,y}$ both even integers:
\begin{equation}
M[\u]=M[\u+2\hat{\x}]=M[\u+2\hat{\bm{y}}], 
\end{equation}
which leads to the Fourier expansion
\begin{equation}
M[\u]=\sum_{\k \in (\pi \mathbb{Z}, \pi \mathbb{Z} )} e^{i\k\cdot\u}M_{\k}.
\label{eq:Fourier}
\end{equation}
Now, we consider the invariance of Eq. \eqref{eq:H_1} under the simultaneous translation of both layers by one lattice spacing. For example, under the $T_x$ of both layers, we have
$\psi_b^\dagger M[\u]\psi_t +h.c. \rightarrow
\psi_b^t e^{-2i\K\cdot\u_b}\tau^1\sigma^2M[\u]e^{2i\K\cdot\u_t}\sigma^2\tau^1 \psi_t^* + h.c.$. The invariance of Eq. \eqref{eq:H_1} requires the two expressions before and after the $T_x$ transformation to be identical, namely 
$M[\u]=-e^{-2i\K\cdot\u}\tau^1\sigma^2M^*[\u]\sigma^2\tau^1.$
In terms of the Fourier components of $M[\u]$, the requirement imposed by $T_x$ can be written as 
\begin{equation}
T_x:\quad M_{-\k-2\K}=-\tau^1\sigma^2M_{\k}^*\sigma^2\tau^1.
\label{eq:Tx_Fourier}
\end{equation}
Similarly, the interlayer tunneling Eq. \eqref{eq:H_1} should also be invariant when both layers are simultaneously acted on by the symmetry actions $T_y$, $\mathcal{M}_x$, $\Rot$ and $\mathcal{T}$. We can summarize all the symmetry constraints on the Fourier components of $M[\u]$ as:
\begin{equation}
\begin{split}
& T_x:\quad M_{-\k-2\K}=-\tau^1\sigma^2M_{\k}^*\sigma^2\tau^1, \\
& T_y:\quad M_{-\k-2\K}=-\tau^1\mu^3\sigma^2M_{\k}^*\sigma^2\mu^3\tau^1, \\
& \mathcal{M}_x:\quad M_{\mathcal{M}_x\k + \K'-\K}=W_1^\dagger M_{\k} W_1,\\
& \Rot:\quad M_{\Rot \k+ \K'-\K}=W_2^\dagger M_{\k} W_2, \\
& \mathcal{T}:\quad M_{\k}=-\tau^3\mu^3M_{\k}\mu^3\tau^3.
\end{split}
\label{eq:sym_Fourier}
\end{equation}
Here, remember that $\K' = \mathcal{M}_x \K = \Rot \K$. 
In addition, we impose an extra symmetry $S$ that exchanges the two layers:  $\psi_t\leftrightarrow \psi_b$ and $\u_t \leftrightarrow \u_b$ (or equivalently $\u\leftrightarrow -\u$). The invariance of the interlayer tunneling under $S$ leads to the additional constraint
\begin{equation}
S:\quad M_{\k}=M_{\k}^\dagger.
\label{eq:exchange}
\end{equation}
The set of conditions in  Eq. \eqref{eq:sym_Fourier} and Eq. \eqref{eq:exchange} relate the Fourier component $M_{\k}$ of $M[\u]$ with other Fourier components within the set $\{M_{\k'} \, | \, \k' = \mathcal{M}_x^m \Rot^n(\k+\K) - \K ,m=0,1, n=0,1,2,3\}$ which contains either four or eight elements depending on the momentum $\k$. In general, we expect that the Fourier components $M_{\k}$ decay rapidly for large $|\k|$. As exemplified by the twisted bilayer graphene case, it should suffice to take the minimal set of Fourier components that contains the smallest allowed $\k$ and other symmetry-related Fourier components. (The general form of higher-momentum Fourier components $M_{\k}$ is discussed in App. \ref{app:general_tunneling}.) In our case, the minimal set is given by  $\{ M_{\k = 0}, M_{ -2\K }, M_{\K' -\K}, M_{-\K' -\K} \}$ with $\k = 0$. The symmetry constraints require that \begin{equation}
\begin{split}
& M_{\k=0} = - M_{-2\K}=w_1\tau^1+w_2\mu^3\tau^1 \equiv M_0,  \\
& M_{\K'-\K}= -M_{-\K'-\K} = w_1\tau^2-w_2\mu^3\tau^2\equiv M_1, 
\end{split}
\label{eq:M_min}
\end{equation}
where the coupling constants $w_1$ and $w_2$ are both real numbers. Within the minimal set of Fourier components, the general form of interlayer tunneling is specified by 
\begin{equation}
M[\u]=2i e^{-i\K\cdot\u}[M_0\sin(\K\cdot\u)+M_{1}\sin(\K'\cdot\u)]
\label{eq:M_u}
\end{equation}
and the full continuum Hamiltonian of the deformed bilayer staggered-flux square lattice is given by Eq. \eqref{eq:H_x}, Eq. \eqref{eq:H_1} and Eq. \eqref{eq:M_u}.

From Eq. \eqref{eq:M_u}, it is interesting to notice that in the case where two un-deformed square-lattice layers are stacked on top of each other with no relative displacement, i.e. $\u=0$, the interlayer continuum Hamiltonian vanishes. In fact, this statement on the vanishing of $M[\u]$ at $\u  = 0$ is not just restricted to the minimal set of Fourier components (see App. \ref{app:general_tunneling} for more detail). It is satisfied even without any truncation in the Fourier components of $M[\u]$. However, we would like to point out that $M[\u=0] = 0$ does not imply the vanishing of interlayer tunneling at the lattice scale. Rather, it means that the lattice-scale interlayer tunneling, if exist, can at most lead to subdominant terms such as terms with derivatives of $\psi_{1,2}$ in the continuum theory. Here, we have made an implicit assumption that the lattice-scale interlayer tunneling is weak compared to the energy scale of the hopping within each layer. We will neglect the subdominant terms in the interlayer tunneling.

When  $\u=0$, $\u=\pm \hat{\bm{x}}$ or $\u=\pm \hat{\bm{y}}$, the sites of the top layer are directly on top of the sites of the bottom layer. In the general setting where $\u$ depends on the spatial location, the factors $\sin(\K\cdot\u)$ and $\sin(\K'\cdot\u)$ of the inter-layer tunneling Eq. \eqref{eq:M_u} suggest that the most contribution to the interlayer tunneling comes from the regimes where $\u$ is locally close to $\pm \hat{\bm{x}}$ or $\pm \hat{\bm{y}}$. In Sec. \ref{subsec:uniform}, we will discuss the case with a uniform deformation where $\u = \hat{\bm{x}}$ is constant in space. The physical meaning of the parameters $w_{1,2}$ will become clear in this discussion. Other cases with $\u= -\hat{\x}$ or $\pm \hat{\bm{y}}$ is similar to case of $\u= \hat{\x}$.

\subsection{Uniform deformation}
\label{subsec:uniform}

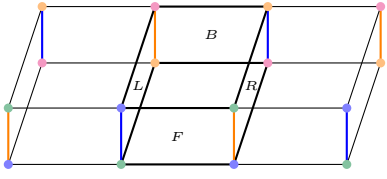
\begin{figure}[htbp]
    \centering
    \begin{tikzpicture}[scale=1.5]
    \draw (1,0)--(0,0)--(0.3,0.9)--(1.3,0.9);
    \draw (1,0.5)--(0,0.5)--(0.3,0.9+0.5)--(1.3,0.9+0.5);
    \draw[thick] (1.3,0.9)--(1,0)--(2,0);
    \draw[orange,thick] (1.3,0.9)--(1.3,0.9+0.5);
    \draw[orange,thick] (3.3,0.9)--(3.3,0.9+0.5);
    \draw[thick] (1.3,0.9+0.5)--(1,0.5)--(2,0.5);
    \draw[blue,thick] (1,0.5)--(1,0);
    \draw[blue,thick] (3,0.5)--(3,0);
    \draw[thick] (2,0)--(2.3,0.9)--(1.3,0.9);
    \draw[thick,blue] (2.3,0.9)--(2.3,0.9+0.5);
    \draw[thick,blue] (0.3,0.9)--(0.3,0.9+0.5);
    \draw[thick] (1.3,0.9+0.5)--(2.3,0.9+0.5)--(2,0.5);
    \draw[thick,orange] (2,0.5)--(2,0);
    \draw[thick,orange] (0,0.5)--(0,0);
    \draw (2,0)--(3,0)--(3.3,0.9)--(2.3,0.9);
    \draw (2,0.5)--(3,0.5)--(3.3,0.9+0.5)--(2.3,0.9+0.5);
    \fill[blue!50] (0,0) circle (0.04);
    \fill[blue!50] (2,0) circle (0.04);
    \fill[blue!50] (1,0.5) circle (0.04);
    \fill[blue!50] (3,0.5) circle (0.04);
    \fill[ForestGreen!50] (1,0) circle (0.04);
    \fill[ForestGreen!50] (3,0) circle (0.04);
    \fill[ForestGreen!50] (0,0.5) circle (0.04);
    \fill[ForestGreen!50] (2,0.5) circle (0.04);
    \fill[orange!50] (1.3,0.9) circle (0.04);
    \fill[orange!50] (3.3,0.9) circle (0.04);
    \fill[orange!50] (2.3,0.9+0.5) circle (0.04);
    \fill[orange!50] (0.3,0.9+0.5) circle (0.04);
    \fill[magenta!50] (0.3,0.9) circle (0.04);
    \fill[magenta!50] (2.3,0.9) circle (0.04);
    \fill[magenta!50] (3.3,0.9+0.5) circle (0.04);
    \fill[magenta!50] (1.3,0.9+0.5) circle (0.04);
    \node at (1.5, 0.25) {\tiny $F$};
    \node at (1.8, 0.25+0.9) {\tiny $B$};
    \node at (1.15, 0.7) {\tiny $L$};
    \node at (2.15, 0.7) {\tiny $R$};
    \end{tikzpicture}
    \caption{The bilayer staggerd-flux square lattice with a uniform relative deformation $(u_x,u_y)=(1,0)$ is depicted. The blue, green, orange, and pink vertices correspond respectively to the lattices site indexed by $i=1,2,3,4$ following Eq. \ref{eq:sublattice}. 
    In the horizontal plane, the unit cell of the bilayer system still covers two lattice spacings in both the $x$- and $y$-directions. There are 4 vertical plaquettes in each unit cell labeled as $F$, $B$, $L$ and $R$. There are 4 vertical links in each unit cell colored as blue, green, orange and pink. 
    The interlayer tunneling only occurs along vertical links. See the main text for the tunneling strength along each link. The interlayer tunneling corresponds to having magnetic fluxes $2\Phi+2 {\phi}-\pi$, $-2\Phi-2 {\phi}+\pi$, $\pi-2\Phi+2 {\phi}$, $-\pi+2\Phi-2 {\phi}$ through the $F$, $B$, $L$, and $R$ plaquettes respectively.
    }
    \label{fig:stack}
\end{figure}

To gain some intuition of the interlayer tunneling term Eq. \eqref{eq:M_u}, we will first discuss the case in which the top layer is rigidly shifted along the $\x$-direction by one lattice spacing while the bottom layer is intact. In this case, the relative deformation $\u$ is uniform in space and is given by $\u= \hat{\x}$, namely $(u_x,u_y)=(1,0)$ in the un-rotated coordinates (which can be also written as $\u=(u_1,u_2)=(1,-1)/\sqrt{2}$ in the $45^\circ$-rotated coordinates). Rewritten in terms of the un-deformed lattice positions $\R$, the Lagrangian coordinates, the interlayer tunneling term is given by 
\begin{equation}
H_1= 2 \int d^2\R~ \psi_b^\dagger(\R)(M_0-M_1)\psi_t(\R+\u)+h.c..
\end{equation}
Transforming it back to the lattice fermions, 
the interlayer Hamiltonian can be understood via Fig. \ref{fig:stack}. Within the $2\times 2$ unit cell of each layer, the four sites, which are labeled by $i=1,2,3,4$ in Eq. \eqref{eq:sublattice}, are colored as blue, green, orange, and pink vertices respectively. The spin-independent interlayer tunneling occurs between sites in the two layers that are connected by vertical links. Depending on the color of the link shown in Fig. \ref{fig:stack}, the hopping terms are different. The amplitudes of the tunneling terms from the top to the bottom layer are given by (1) $2w e^{-i {\phi}}$ for the blue links and (2) $2w e^{i (\phi-\pi)}$ for the orange links. Here $w$ is given by $w=|w_1 + i w_2|$, and $ {\phi}=\arctan(w_2/w_1) + \pi/4$. 

One can understand the interlayer tunneling terms as follows. The unit cell of this bilayer system in the horizontal plane is still $2\times 2$ in lattice spacings. Each such unit cell contains 4 vertical plaquettes labeled as $F$, $B$, $L$, and $R$ respectively as shown in Fig. \ref{fig:stack}. The interlayer tunneling terms described above correspond to having magnetic fluxes $2\Phi+2 {\phi}-\pi$, $-2\Phi-2 {\phi}+\pi$, $\pi-2\Phi+2 {\phi}$, $-\pi+2\Phi-2 {\phi}$ through the $F$, $B$, $L$, and $R$ vertical plaquettes.

In this bilayer system, when the interlayer tunneling is turned off, i.e. $w_1= w_2 =0$, the band structure is gapless only at $\q = (0,0)$. The continuum theory equivalently describes 8 copies of 2-component Dirac fermions. These 8 copies come from two-fold valley, two-fold spin and two-fold layer degrees of freedom. As we turn on a finite interlayer tunneling $w_1$ and $w_2$, one observes that the original gapless point at $\q=(0,0)$ splits into four gapless Dirac points located at  $(q_1,q_2)=\pm (w_1+w_2,w_2-w_1)$, $\pm(w_1-w_2,-w_1-w_2)$ in the $45^\circ$-rotated coordinates, see Fig. \ref{fig_uniform} below. Each of these Dirac points is described by 2 copies of 2-component Dirac fermion with the two copies coming from the two-fold spin degrees of freedom. 
\begin{figure}[htbp]
\centering
\includegraphics[scale=0.5]{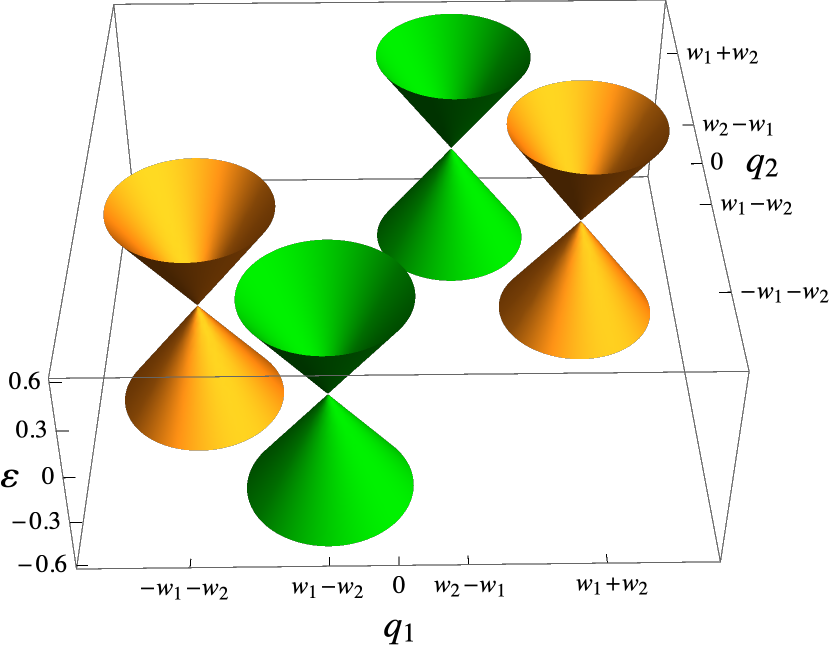}
\caption{The shifted Dirac cones at half-filling for a uniform deformation $\u=\hat{\x}$. The two colors correspond to the two valleys $\mu^3=\pm 1$.}
\label{fig_uniform}
\end{figure}

\subsection{Rigid twist}
\label{subsec:twist}
Now we consider the case of twisted bilayer staggered-flux square lattice where the two layers are deformed by rigid rotations by angles $\pm \theta /2$. We assume the twist angle $\theta$ is small. Hence, the deformation field can be written as $\u_t=-\u_b=\frac{\theta}{2}\hat{\bm{z}}\times\x$, where $\hat{\bm{z}}$ is the unit vector along the ${\bm z}$-direction. The interlayer tunneling can be obtained directly by plugging $\u = \theta \hat{\bm{z}} \times \x$ in Eq. \eqref{eq:M_u}. The Hamiltonian terms within each of the top and the bottom layer are obtained by plugging $\u_{t/b}$ into Eq. \eqref{eq:H_x}. Here, we've also assumed that there is no chemical potential difference between the top and bottom layers. We note that the terms $\psi^\dagger_{t} \tau^j(\partial_j u^i_{t})\partial_i \psi_{t}$ and $\psi^\dagger_{b} \tau^j(\partial_j u^i_{b})\partial_i \psi_{b}$ that captures the rotation of the Dirac cones in each layer each contains two derivatives and, hence, is parametrically small compared to other terms for small twist angle $\theta$ as the relevant physics happens at the moir\'e lattice length scale. We  drop these two terms to simplify the Hamiltonian. One can further simplify the Hamiltonian by the redefining the fields $\psi_l(\x)\rightarrow e^{i\K\cdot\u_l}\psi_l(\x)$ for the two layers $l = t, b$. The continuum model Hamiltonian for the twisted bilayer staggered-flux square lattice now reads 
\begin{equation}
 \resizebox{1\hsize}{!}{$\begin{aligned}
 & H= \int d^2\x~ \Big\{ \sum_{l=t,b} \psi_l^\dagger (\x) ( -i\tau^i\partial_i) \psi_l(\x)  ~\\
 &+ \left( 2i  \psi_b^\dagger(\x) [M_0\sin(Ku_1)+M_1\sin(Ku_2)] \psi_t(\x)+h.c. \right)\Big\},
\end{aligned}$}
 \label{eq:Ham_twist_pre}
\end{equation}
where $K=|\K|=\pi/\sqrt{2}$ and $(u_1,u_2)=(-\theta x_2 , \theta x_1)$ in the $45^\circ$-rotated coordinate. If we did not drop the terms $\psi^\dagger_{t} \tau^j(\partial_j u^i_{t})\partial_i \psi_{t}$ and $\psi^\dagger_{b} \tau^j(\partial_j u^i_{b})\partial_i \psi_{b}$ for each layer, we would need to do an extra layer-dependent rotation of the Dirac spinor 
$\psi_{t/b} \rightarrow e^{\mp i\tau^3 \theta/4} \psi_{t/b}$ to remove these terms from the Hamiltonian. Such a layer-dependent rotation in fact leaves the interlayer tunneling terms invariant. Hence, the Hamiltonian Eq. \ref{eq:Ham_twist_pre} is still valid even when the effect of the intralayer terms $\psi^\dagger_{t} \tau^j(\partial_j u^i_{t})\partial_i \psi_{t}$ and $\psi^\dagger_{b} \tau^j(\partial_j u^i_{b})\partial_i \psi_{b}$ are considered. Now using the identity $\K\cdot\u=\Q\cdot\x$ with $\Q \equiv \theta
(\K\times\hat{\bm{z}})$, $M[\u]$ can be further written as $M(\x)$. Moreover, we turn to the dimensionless parametrization $\tilde{\q}=\q/|\Q|$, $\tilde{\x}=|\Q|\x$ and $\tilde{w}_i=w_i/|\Q|$. Combining all these and omitting the tildes from now on, we arrive at
\begin{equation}
 \resizebox{0.95\hsize}{!}{$\begin{aligned}
& H= \int d^2\x \Big\{ \sum_{l=t,b}  \psi_l^\dagger (\x) (-i\tau^i\partial_i) \psi_l(\x) ~\\
& + \left( 2i \psi_b^\dagger(\x) (-M_0\sin x_2+M_1\sin x_1) \psi_t(\x)+h.c. \right)\Big\},
\end{aligned}$}
\label{eq:Ham_twist}
\end{equation}
Here we have also rescaled the energy by an overall of multiplicative factor. Notice from Eq.  \eqref{eq:M_min} that the two valleys (corresponding to $\mu^3=\pm 1$) are decoupled. In the following, we will only focus on the $\mu^3=+ 1$ valley as the spectrum of the $\mu^3=-1$ valley follows straightforwardly by replacing $w_2\rightarrow -w_2$. 
Focusing on the $\mu^3=+ 1$ valley, we can effectively write $M_0=(w_1+w_2)\tau^1\equiv w_s \tau^1$ and $M_1=(w_1-w_2)\tau^2\equiv w_a\tau^2$. We will present the analytical study of the Hamiltonian Eq. \eqref{eq:Ham_twist} in the next section. Before that, we comment on two special limits: (1) When $w_1=\pm w_2$, we are in the ``chiral limit'' where either $M_0$ or $M_1$ vanishes and the interlayer interaction becomes uniform in either the $x_1$- or $x_2$-direction. In this case, the moir\'e superlattice is quasi-one-dimensional. (2) When $w_2=0$, the Hamiltonian is independent of the valley Pauli matrix $\bm{\mu}$ and recovers the $\SU(4)$ symmetry (acting on the valley- and spin- spaces) of the decoupled-layer case.

\begin{figure}[htbp]
\centering
\includegraphics[scale=0.25]{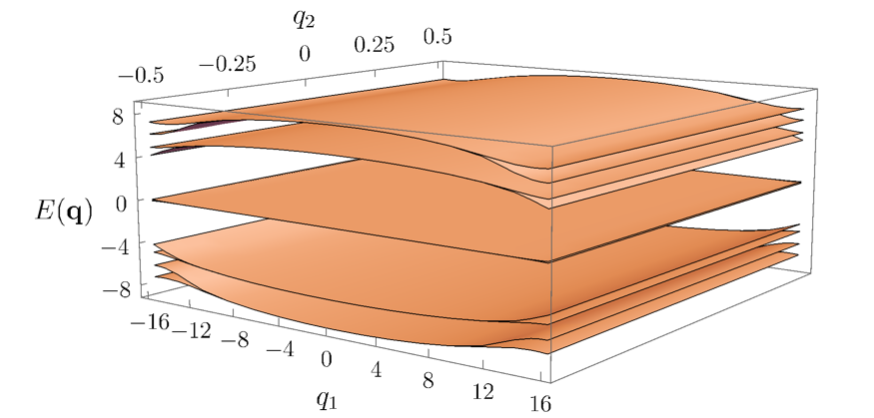}\\
\vskip 0.3cm
\includegraphics[scale=0.25,trim=0 0 0.1cm 0.3cm,clip]{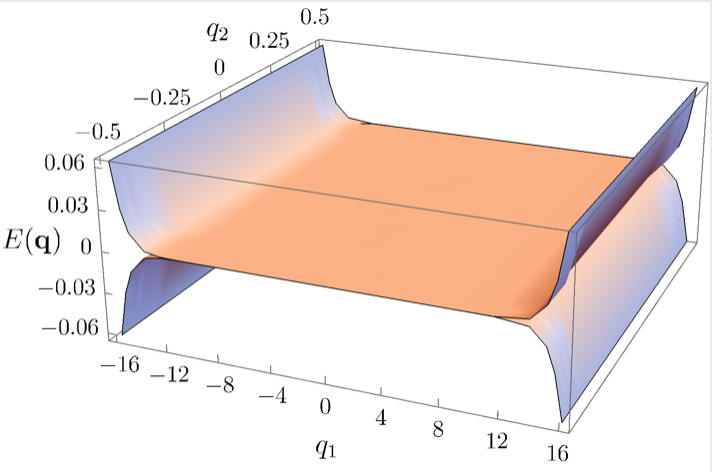}
\caption{Top panel: Band structures in the chiral limit $w_s=10$,  $w_a=0$. The momentum $q_1$ is conserved while $q_2$ takes value in the moir\'e Brillouin zone $q_2\in (-1/2,1/2)$. 
Bottom panel: Zoomed-in view of the flattened bands (with the coupling constants $w_a$ and $w_s$) near half filling.}
\label{fig:flat}
\end{figure}

We can numerically compute the spectrum of the Hamiltonian \eqref{eq:Ham_twist}. We start by discussing the chiral limit. In the chiral limit, the system has continuous translation symmetry along either the $x_1$- or the $x_2$-direction. For example, when $w_1 = w_2$, i.e. $w_a = 0$, the momentum $q_1$ along the $x_1$-direction is conserved, while the momentum $q_2$, still being the crystal momentum along the $x_2$-direction, is conserved only modulo integer and, hence, has a Brillouin zone of $[-1/2,1/2)$. 
In Fig. \ref{fig:flat}, we show an example of the band structure in the chiral limit $w_a=0$ and $w_s =10$. Both the top and bottom panels correspond to the same parameters. In this band structure for which the valley index $\mu^3 = +1$ is already fixed, every band shown in Fig. \ref{fig:flat} has a two-fold degeneracy (in addition to the two-fold spin degeneracy). From the top panel of Fig. \ref{fig:flat}, we notice that the two bands near the half filling are extremely flat and close to zero energy in a large area in the momentum space. The bottom panel of Fig. \ref{fig:flat} shows the zoomed-in view of the two bands near half-filling showing they are still dispersive bands whose energies are not exactly zero. The deviation from zero energy grows as the momentum $q_1$ increases. Here, we emphasize that there is in fact no gap separating the two bands near half filling from other bands. These two bands near half filling will overlap with other bands in energy for larger values of $q_1$ that is beyond the range plotted in Fig. \ref{fig:flat}.  We will discuss the analytical understanding of the two-fold degeneracy and the emergence of a large momentum-space region where the bands flatten in the next section. Also, we will show that the the entire spectrum must be free of gaps at any energy.

\begin{figure}[htbp]
\centering
\includegraphics[scale=0.12,trim=0 0.3cm 0 0.5cm, clip]{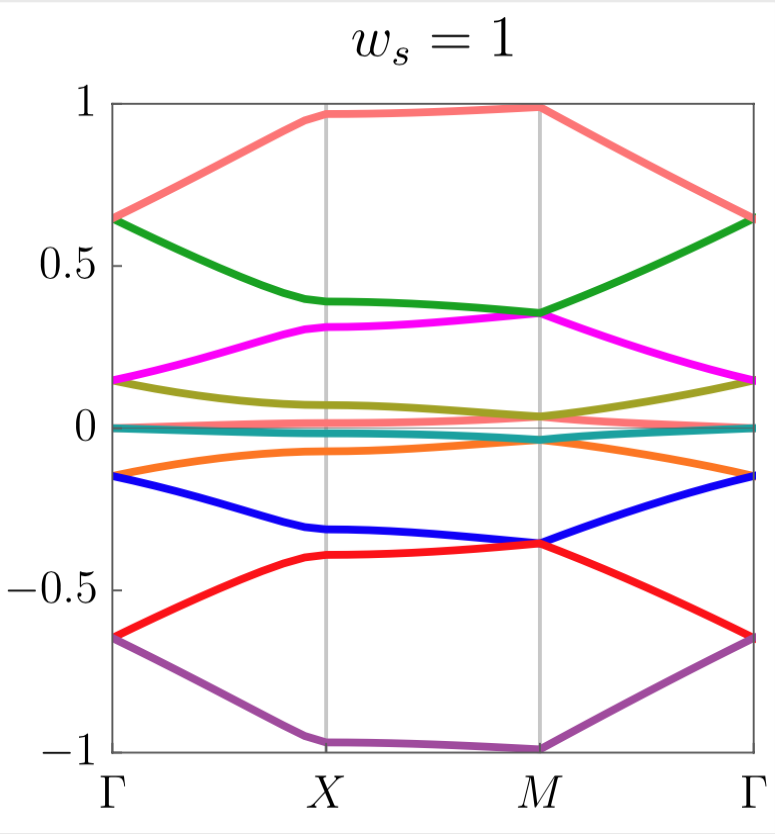}
\includegraphics[scale=0.12,trim=0 0.3cm 0 0.5cm, clip]{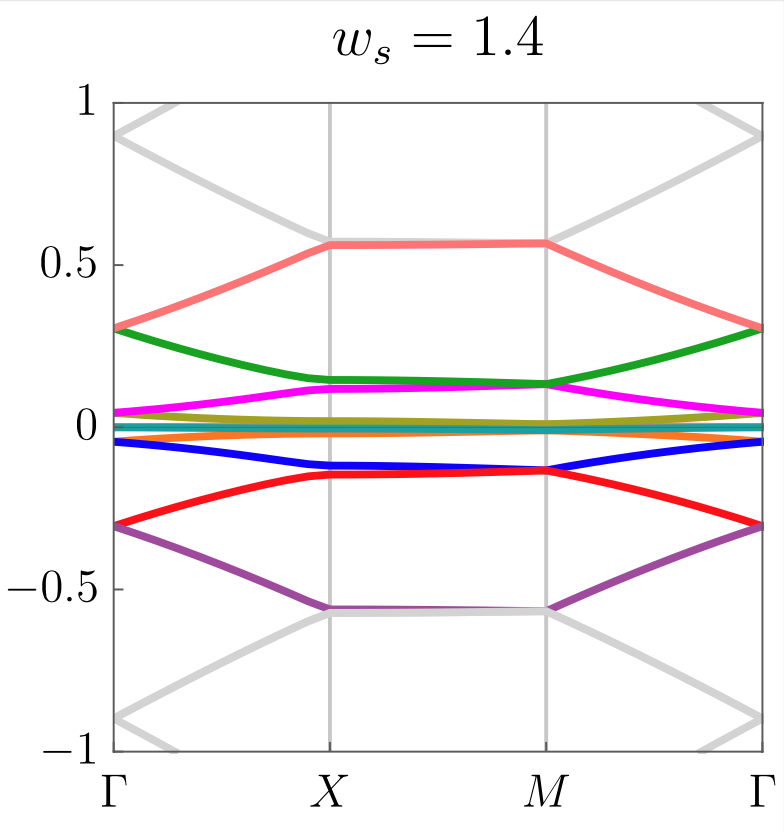}\\
\vskip 0.2cm
\includegraphics[scale=0.12,trim=0 0.3cm 0 0.5cm, clip]{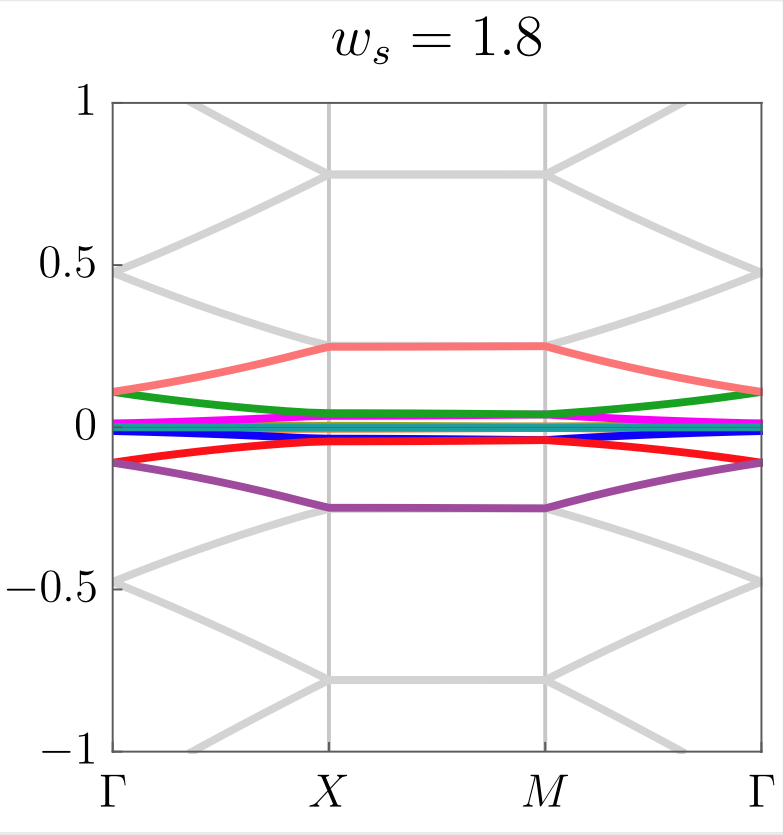}
\includegraphics[scale=0.129,trim=0 0.3cm 0.3cm 0.5cm, clip]{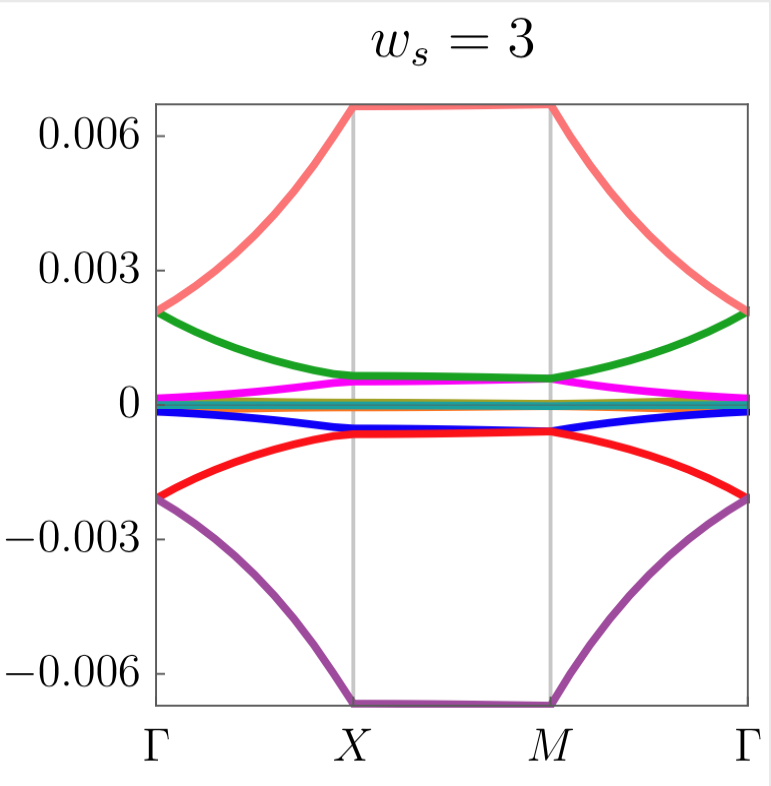}
\caption{Band structures along the moir\'e Brillouin zone trajectory $\Gamma \rightarrow X \rightarrow M \rightarrow \Gamma$ for various $w_s$'s with a fixed $w_a=0.5$ is plotted. For fixed spin and valley indices, the spectrum is two-fold degenerate (corresponding the quantum number $s=\pm 1$ in Eq. \eqref{eq:ansatz}). All the bands are connected by Dirac cones at the $\Gamma$ and the $M$ points in the moir\'e Brillouin zone. The ten bands (with each doubly degenerate)
which are nearest to the zero energy are colored. As $w_s$ increases, the spectrum are compressed towards the zero energy and the band near zero-energy are flattened. 
Note that the energy range of the plot with $w_s=3$ in the right bottom panel is about 200 times smaller than the other plots.}
\label{fig:nonchiral}
\end{figure}

When we move away from the chiral limit, both momenta $q_{1,2}$ become crystal momenta defined within the moir\'e Brillouin zone i.e. $q_{1,2} \in [-\pi, \pi)$. Fig. \ref{fig:nonchiral} shows the single-valley band structures (with $\mu^3 = +1$) for various $w_s = 1, 1.4, 1.8, 3$ and a fixed $w_a =0.5$. The plotted band structures are the band structure along a momentum-space contour that connects the $\Gamma$ point $(q_1, q_2) = (0,0)$, the $X$ point $(q_1, q_2) = (\pi,0)$, and the $M$ point $(q_1, q_2) = (\pi,\pi)$. Similar to the chiral limit, each band is two-fold degenerate in additional to the two-fold spin degeneracy. In Fig. \ref{fig:nonchiral}, the colored bands are the ten bands that are closest to zero energy. 
We see that, as $w_s$ increases, the spectrum gets compressed towards zero energy and the bands near zero energy flattened, which yields a large number of low-energy states. We also notice that each band is connected with its neighboring bands via Dirac cones at the $\Gamma$ point or at the $M$ point. Hence, the entire band structure is ``infinitely connected" and is free of band gaps at all energies. We will discuss the two-fold degeneracy of each band, the flattening of the bands near zero energy and the infinite connectivity of the bands in the next section. We will also discuss the analytical understanding of the infinite 
connectivity of the bands shown in Fig. \ref{fig:nonchiral}.

\section{Analytical study of the Magic continuum}
\label{sec:magic}
In this section, we discuss the magic continuum of the twisted bilayer staggered-flux square lattice where the two layers are deformed by rigid rotations of angles $\pm \theta /2$ respectively. As discussed in Sec. \ref{subsec:twist}, upon the field redefinition and the re-scaling of momentum and energy, the Hamiltonian of this twisted bilayer is given by Eq. \eqref{eq:Ham_twist}. As explained above, it suffices to focus only on the $\mu^3 = +1$ valley. We notice that the problem can be further simplified using the following basis of the single-particle wavefunction 
\begin{equation}
\left(\begin{array}{c}
     \psi_t  \\
     \psi_b 
\end{array} \right) = \frac{1}{\sqrt{2}} \left( \begin{array}{c}
     \psi  \\
     i s \psi 
\end{array} \right)
\label{eq:ansatz}
\end{equation}
The Hamiltonian Eq. \eqref{eq:Ham_twist} of the twisted bilayer system is diagonal in the $s = \pm 1 $ basis. We would like to comment that, as shown in App. \ref{app:general_tunneling}, $s$ is always a good quantum number in the twisted bilayer system even when we consider the most general form of spin-independent interlayer tunneling $M[\u]$ (including the Fourier components beyond the minimal set) allowed by the constraints discussed in Sec. \ref{subsec:bootstrap}. In the following discussions, we will still focus on the Hamiltonian Eq. \eqref{eq:Ham_twist} where the interlayer tunneling involves the minimal and most dominant set of Fourier components. Physically, the quantum number $s$ can be understood as inherited from the spatial regions where $\u$ is locally close to $\u =\pm 
\hat{\x}$ and $\u =\pm 
\hat{\bm{y}}$ and where the interlayer tunneling acquires its most contribution from.  In these spatial regions, each of combination of the quantum number $s=\pm 1$ and the valley index $\mu^3=\pm 1$ is locally associated with one of the four Dirac cones shown in Fig. \ref{fig_uniform} (obtained with the uniform deformation). For a given quantum number $s$ (and the fixed valley index $\mu^3 =1$), the single-particle Hamiltonian that acts on $\psi$ reads 
\begin{equation}
\begin{split}
h = \tau^1(-i\partial_1-2sw_s\sin x_2)
 +\tau^2(-i\partial_2+2sw_a\sin x_1).
\end{split}
\label{eq:period_B}
\end{equation}
This Hamiltonian equivalently describes a Dirac fermion in a periodically modulated effective background magnetic field
$B(\x)=2s(w_a\cos x_1+w_s\cos x_2)$ written in a Coulomb gauge. 
The problem of Dirac fermion in an effective periodic magnetic field has been shown to emerge and has been investigated in the contexts of strained graphene \cite{guinea2008midgap, wehling2008midgap}, graphene in a field \cite{snyman2009gapped,tan2010graphene} and strained topological crystalline insulators \cite{Liang}. As a brief remark, our analysis has been focusing on the limit $t/\Delta = 1$ where the continuum description of the single-layer theory Eq. \eqref{eq:Ham} has an isotropic Dirac velocity. When we take $t/\Delta\neq 1$ in Eq. \eqref{eq:lattice_ham}, namely when the Dirac velocity of the single-layer theory develops a valley-dependent anisotropy, 
the above equation \eqref{eq:period_B} would only get modified by a valley-dependent velocity anisotropy. The terms induced by the interlayer tunneling remain intact and all the analyses below can still carry over. Hence, we will continue the analysis in the isotropic limit in the following.

In this section, we will discuss the exponential reduction of the Dirac velocity in the Hamiltonian Eq. \eqref{eq:period_B}, the emergence of flattened bands and the associated large number of low-energy states in the magic continuum.  We will also discuss the analytical understanding of the infinite connectivity of all the bands shown in Fig. \ref{fig:nonchiral}.

\subsection{Exact zero-energy states at the Dirac point and the renormalized Dirac velocity}

Similar to the twisted bilayer graphene system, the band structure of the twisted bilayer staggered-flux square lattice contains Dirac cones near half filling inherited from each of the staggered-flux square-lattice layer. The location of the inherited Dirac points in the twisted bilayer system should be at $(q_1, q_2) = (0,0)$ for the Hamiltonian Eq. \eqref{eq:period_B}. 

The zero-energy eigenstates exactly at the Dirac point can be solved analytically. Notice that the zero-energy eigenstates of $h$ in Eq. \eqref{eq:period_B} should also be the eigenstates of $\tau^3$. The analytical expression of the exact zero-energy eigenstates are given as 
\begin{equation}
\psi_+ = c_+ \left(\begin{matrix} e^{- B(\x)}\\ 0 \end{matrix}\right),\quad \psi_- = c_- \left(\begin{matrix} 0 \\ e^{B(\x)} \end{matrix}\right),
\label{eq:sol}
\end{equation}
where $c_\pm$ are the normalization constants which can be fixed by integrating over the moir\'e unit cell, 
\begin{equation}
c_+=c_- = [4\pi^2 I_0(4w_s)I_0(4w_a)]^{-1/2}.
\label{eq:norm}
\end{equation}
Here $I_0$ is the modified Bessel function of the first kind. Notice that the exact zero-energy eigenstates $\psi_\pm$ satisfy the periodic condition that
\begin{equation}
\psi_\pm (x_1,x_2)=\psi_\pm (x_1+2\pi,x_2)=\psi_\pm (x_1,x_2+2\pi),
\end{equation}
which is in agreement with the expectation that the Dirac point of the bilayer system is located at $(q_1,q_2) = (0,0)$ within the moir\'e Brillouin zone. One can also prove that these solutions are unique. 

We now compute the Dirac velocity. In the momentum space, $\psi_\pm$ are the eigenstates solutions at the Dirac point $(q_1, q_2) = 0$. The effective single-particle Hamiltonian for small $q$ near $\q = (0,0)$ can be obtained by treating the term $q_i \tau^i$ as a perturbation to the subspace formed by the solutions $\psi_\pm$ at $q=0$. The matrix elements of the perturbation $q_i \tau^i$ are given by
\begin{equation}
h_{\q,\xi\xi'}=\int_{u.c.} d^2\x~ \psi_{\xi}^\dagger (\x) (\tau^i q_i) \psi_{\xi'}(\x),
\end{equation}
where $\xi, \xi'=\pm$ and $\int_{u.c.} d^2\x$ denotes the integration over a moir\'e unit cell (in the real space). Here $\psi_{\xi'}$ and $\psi_{\xi'}^\dag$ represents the wavefunction $\psi_{\pm}(\x)$ and their conjugate, which should not be confused with fermion operators. Plugging in the expressions Eq. \eqref{eq:sol} and Eq. \eqref{eq:norm}, we arrive at
\begin{equation}
h_{\q}=\tau^iq_i/I_0(4w_a)I_0(4w_s),
\label{eq:Dirac_VReduced}
\end{equation}
which captures the dispersion of the bands near zero energy in the vicinity of $\q = (0,0)$. Eq. \eqref{eq:Dirac_VReduced} describes a gapless Dirac cone with a renormalized Dirac velocity. 
A similar analysis of the renormalization of the Dirac velocity was also given in Ref. \onlinecite{snyman2009gapped} which studied a monolayer graphene under general periodic magnetic and electric fields.

From Eq. \eqref{eq:Dirac_VReduced}, the renormalized Dirac velocity is given by
\begin{align}
v^*=1/I_0(4w_a)I_0(4w_s).
\label{eq:renormalized_velocity}
\end{align}
We are interested in how $v^*$ changes as the coupling constants $w_a$ and $w_s$ vary.
The function $I_0(z)$ satisfies $I_0(0)=1$ and increases monotonically and exponentially with $|z|$. When $w_s=w_a=0$, $v^*=1$ recovers the results for the decoupled bilayer. As $|w_s|$  and/or $|w_a|$ increase, the Dirac velocity $v^*$ becomes exponentially suppressed. At the same time, the bands near half filling are flattened, leading to a large number near-zero-energy states. This exponential suppression of the Dirac velocity and the emergence of flattened bands occur in a very large range of $w_a$ and $w_s$, which is in contrast to the twisted bilayer graphene system where the drastic reduction of the Dirac velocity and bandwidth only occurs around a discrete set of coupling constants and twist angles. Hence, there is a magic continuum in the twisted bilayer staggered-flux square lattice system.  

\begin{figure}
\includegraphics[scale=0.17,trim=0 0 0 0.09cm, clip]{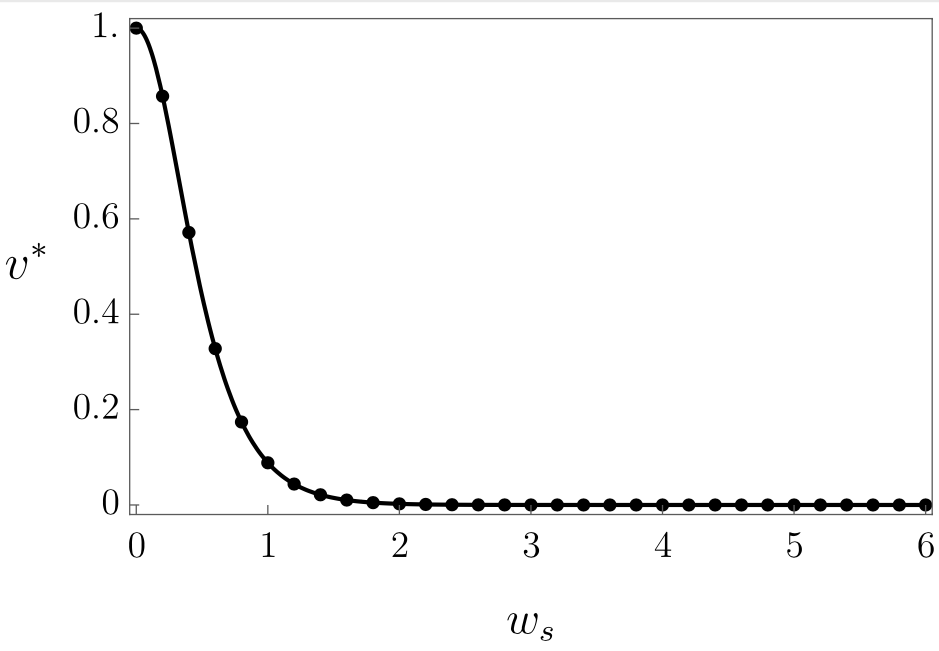}
\caption{The renormalized Dirac velocity $v^*$ as a function of $w_s$ in the chiral limit $w_a=0$. There is a magic continuum for $w_s\gtrsim 1$. The dots are the numerically computed Dirac velocity. 
The smooth line is the modified Bessel function $I_0(4w_s)^{-1}$, which is the analytical expression of the reduced Dirac velocity $v^*$ we obtained in Eq. \eqref{eq:renormalized_velocity}. }
\label{fig:magic}
\end{figure}

In Fig. \ref{fig:magic}, we plot the Dirac velocity $v^*$ as a function of $w_s$ obtained from numerically computing the spectrum, which shows a perfect match with the analytical expression above. For simplicity, we've only focused on the chiral limit $w_a=0$ where $q_1$ is conserved in Fig. \ref{fig:magic}. The expression of the renormalized Dirac velocity Eq. \eqref{eq:renormalized_velocity} is generally applicable for any parameters $w_a$ and $w_s$.

\subsection{Intuitive understanding of the emergence of a large number of low-energy bands}

In this subsection, we provide an intuitive understanding of the emergence of a large number of low-energy bands as $w_a$ and $w_s$ increase. Here, by low-energy bands, we refer to the bands with energies close to zero, namely close to half filling.
Recall that in a uniform magnetic field of strength $B$, the massless Dirac fermions form a set of Landau levels. Each quantum state occupies an area of $2\pi l_B^2$ with $l_B=1/\sqrt{B}$ the magnetic length, and the degeneracy of each Landau level is the ratio of the full area of the system divided by $2\pi l_B^2$. For a spatial region with a finite size, the Landau-level degeneracy can be well-approximated by the number of magnetic flux quanta contained in this region (regardless of the sign of the magnetic field). For a review, see for example [\onlinecite{goerbig2009quantum}]. 

When the magnetic field is slowly varying over $l_B$, the Landau levels remain a good approximation. But the Landau levels will have position-dependent energies
\begin{equation}
E_n(\x)=\pm \sqrt{2n |B(\x)|}, \label{eb}
\end{equation}
where $n\in \mathbb{Z}$ labels the different Landau levels. Because of the dependence on $\x$, Landau levels at different positions are no longer degenerate and collectively form dispersive bands. The $n=0$ Landau level is rather special as its energy does not depend on $B(\x)$. However, the chirality (labeled by the $\pm$ signs in Eq. \eqref{eb}) of Landau level depends on the sign of $B(\x)$. For the twisted bilayer staggered-flux square lattice, the effective magnetic field $B(\x) = 2s(w_a\cos x_1+w_s\cos x_2)$ has a spatially dependent sign. Therefore, different spatial regions will host the $n=0$ Landau level of opposite chiralities, which yields dispersive modes on the interfaces between these regions and also perturbs the states in the local $n=0$ Landau level away from exactly zero energy. Even though the state in the local $n=0$ Landau level will no longer have exactly zero energy, their energies will still remain close to zero when the local magnetic length is much smaller than the length scale of variation of $B(\x)$.
To make an estimate of the number of states with energy close to zero, we first divide a spatial moir\'e unit cell, which can be chosen as $(x_1, x_2) \in [-\pi/2, 3\pi/2) \times [-\pi/2, 3\pi/2)$, into 4 subregions $A_1=[-\pi/2, \pi/2) \times [-\pi/2, \pi/2)$, $A_2=[\pi/2, 3\pi/2) \times [-\pi/2, \pi/2)$, $A_3=[-\pi/2, \pi/2) \times [\pi/2, 3\pi/2)$ and $A_4=[\pi/2, 3\pi/2) \times [\pi/2, 3\pi/2)$, whose centers are the 4 local extrema of $B(\x)$. 
For every subregion, we replace the effective magnetic field $B(\x)$ by its average value in the same subregion. With this replacement, the numbers of magnetic flux quanta through subregions $A_1$ and $A_4$ are both $2|w_s+w_a|=4|w_1|$, while the numbers of flux quanta through $A_2$ and $A_3$ are both $2|w_s-w_a|=4|w_2|$. Their sum is proportional  $|w_1|+ |w_2|$ and gives an estimate for the number of close-to-zero-energy state in a moir\'e unit cell. Therefore, the number of close-to-zero-energy moir\'e bands in the band structure of the Hamiltonian Eq. \eqref{eq:period_B} should be $\propto (|w_1|+ |w_2|)$. 

Note that the estimation above relies on the slow variation of the $B(\x)$ field on the scale of the local magnetic length, which amounts to the requirement that both $|w_1|,|w_2|\gg 1$. If we only have $|w_1|\gg 1$ but $|w_2|$ is small, then the earlier estimations in the $A_2$ and $A_3$ subregions won't be controlled while the arguments for $A_1$ and $A_4$ still work. Therefore the degeneracy in this case is at least proportional to $|w_1|$. Similar arguments follow for the case with $|w_2|\gg 1$ and $|w_1|$ small.

\begin{figure}[htbp]
\centering
\includegraphics[scale=0.13,trim=0 0 0.05cm 0.05cm,clip]{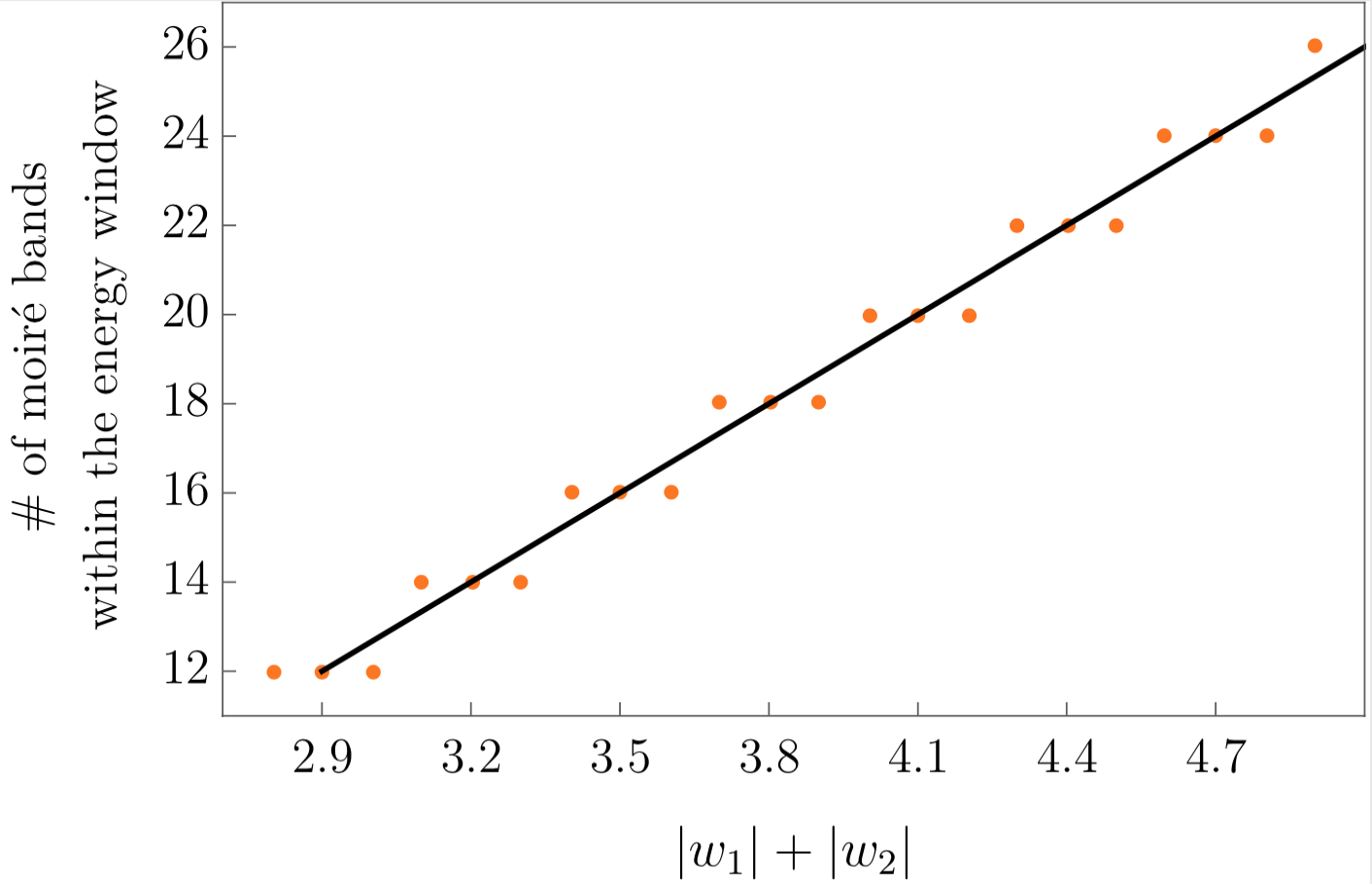}\\
\caption{Number of moir\'e bands with nearly-zero energies scales linearly with $|w_1|+|w_2|$ in the regime with $|w_{1}|, |w_2| \gtrsim 1$. We fix $|w_1|-|w_2|=1$ in the plot and count all the bands (obtained from numerical calculations) that completely lie within the energy window $E\in (-0.05,0.05)$. Note that bands only partially lying inside the energy window are not included in the counting. The step-like feature is due to the fact that the number of moir\'e bands counted has to be an integer.
The distance between steps is always two due to the fact that the spectrum is symmetric in energy about the $E=0$ axis. The black straight line is obtained from connecting the middle points of each step.
}
\label{fig:deg}
\end{figure}

We can compare the analysis above with numerical calculations. This band structure is free of a band gap. In order to numerically estimate the number of low-energy bands in the system, we have to choose a specific small energy window around zero energy and only count the number of (numerically obtained) bands fully contained inside this energy window. In Fig. \ref{fig:deg}, we plot the the numerically calculated number of moir\'e bands fully contained within the small energy window $E\in (-0.05,0.05)$ as a function of $|w_1|+|w_2|$. Fig. \ref{fig:deg} exhibits a linear behavior in the parameter regime $|w_1|,|w_2|\gtrsim 1$, consistent with the previous analysis. If we change the size of the energy window, the number of bands within the energy window remains linearly dependent on $|w_1|$+$|w_2|$ but with a different slope.

\subsection{Infinite Band Connectivity }
The band structure shown in Fig. \ref{fig:nonchiral} is free of band gaps. All the bands are connected to each other. In this subsection, we show that this infinite connectivity of the band structure is demanded by the symmetry of the Hamiltonian Eq. \eqref{eq:period_B}. The first relevant symmetry is the original two-fold spatial rotation symmetry $\Rot^2: x_{1,2} \rightarrow -x_{1,2}\,, ~\psi(\x) \rightarrow -\tau^z \psi(\Rot^2\x)$. For a fixed quantum number $s$ (as well as a fixed valley index $\mu^3=\pm1$ and a fixed spin species), the Hamiltonian Eq. \eqref{eq:period_B} is invariant under a new symmetry action $\RInv\Tt$ that combines a two-fold spatial rotation $\RInv: x_{1,2} \rightarrow -x_{1,2}+\pi \,, ~\psi(\x) \rightarrow \psi(\RInv\x)$ and a time-reversal transformation $\Tt: \psi(\x) \rightarrow \tau^1 \psi(\x),~ i \rightarrow -i $. Note that the symmetries $\RInv$ and $\Tt$ are different from the two-fold spatial rotation symmetry $\Rot^2$ and the time reversal symmetry $\mathcal{T}$ originated from the single-layer model of the staggered-flux square lattice. Each of $\RInv$ and $\Tt$ individually is not a symmetry of the Hamiltonian Eq. \eqref{eq:period_B}, but their combination $\RInv\Tt$ is. The symmetry $\RInv\Tt$ squares to $1$, namely $(\RInv\Tt)^2 =1 $. 

For every band, this symmetry $\RInv\Tt$ ensures that the Berry curvature vanishes at every momentum point where the band structure (with the valley index $\mu^3$, the spin index and the quantum number $s$ all fixed) is non-degenerate.  
At a generic degenerate point ${\bm q}$, we expect the band structure to be locally described by a gapless Dirac cone 
which leads to a $\delta$-function contribution to the Berry curvature with a total flux $\pi$ fully concentrated at the momentum ${\bm q}$. In general, the $\RInv \Tt$ symmetry allows $\delta$-functions in the Berry curvature with fluxes $n \pi $, $n \in \mathbb{Z}$ at a degenerate point. When $|n|>1$, such a degenerate point can be generically split into $|n|$ gapless Dirac cones without breaking the $\RInv \Tt$ symmetry. For an isolated gapless Dirac cone, the symmetry $\RInv\Tt$ forbids a non-zero Dirac mass and, hence, ensures the stability of the gapless Dirac cone. The constraint on the Berry curvature imposed by $\RInv\Tt$ leads to the consequence that each band must contain an even number of gapless Dirac points to ensure that total Berry flux within a band is an integer multiple of $2\pi$. Since the Hamiltonian Eq. \eqref{eq:period_B} also respects the two-fold spatial rotation symmetry $\Rot^2$, gapless Dirac points must come in pairs in the mori\'e Brillouin zone except at the $\Gamma$ point and the $M$ point. In other words, the symmetries $\RInv\Tt$ and $\Rot^2$ together require that the total number of gapless Dirac cones located at the $\Gamma$ point and the $M$ point for each band has to be even.

Now, let's label the bands of the Hamiltonian Eq. \eqref{eq:period_B} by $m= \pm 1, \pm 2, ...$. The band with label $m > 0$ ($m<0$) is the $|m|$-th band above (below) zero energy. Consider starting with vanishing $w_{a,s}$ and gradually turning them on. In the limit where $w_{a,s}$ are zero, the band $m=1$ and the band $m=-1$ are connected via a single gapless Dirac cone centered at the $\Gamma$ point. As we gradually turn on $w_{a,s}$, this gapless Dirac cone between the bands $m=1$ and $m=-1$ is stable and is pinned at zero energy by the time-reversal symmetry $\mathcal{T}$. At the $M$ point, the bands $m=1$ and $m=-1$ are well separated in energy. To ensure that the $m=1$ band contains an even number of gapless Dirac points in total, it has to be connected to the $m=2$ band via a gapless Dirac cone located at the $M$ point. Now, the requirement that the $m=2$ band contains an even number of gapless Dirac points further enforces a Dirac point at the $\Gamma$ point connecting the $m=2$ and the $m=3$ bands. By iterating similar arguments, we can conclude that all the bands of the Hamiltonian Eq. \eqref{eq:period_B} are connected via gapless Dirac cones at the $\Gamma$ point and the $M$ point. 

The arguments above for the infinite connectivity of the bands in Eq. \eqref{eq:period_B} rely on the symmetry $\RInv \Tt$. The Hamiltonian Eq. \eqref{eq:period_B} is obtained from choosing the minimal set of Fourier components in the interlayer tunneling $M[\u]$. In App. \ref{app:general_tunneling}, we show that the symmetry $\RInv \Tt$ is present even when we consider the most general form of spin-independent interlayer tunneling $M[\u]$ allowed by the constraints discussed in Sec. \ref{subsec:bootstrap}. Hence, the infinite connectivity of the bands is present under the general allowed spin-independent interlayer tunneling $M[\u]$. 

\section{Discussion}
\label{sec:discuss}

To test the robustness of the our analysis, one can explicitly break certain microscopic symmetries. For example, when the reflection symmetry $\mathcal{M}_x$ is broken, the terms $w_3\tau^2+w_4\mu^3\tau^2$ can be added to $M_0$ in Eq. \eqref{eq:M_min}, while $-w_3\tau^1+w_4\mu^3\tau^1$ can be added to $M_1$ in Eq. \eqref{eq:M_min}, with $w_3$, $w_4$ both real. The Hamiltonian for the twisted bilayer staggered-flux square lattice system can again be understood as describing a massless Dirac Hamiltonian in the same periodic effective magnetic field, but in a different gauge for the corresponding vector potential. Similar properties are found if the rotation symmetry $\Rot$ is broken while $\Rot^2$ is still preserved. We've also checked that breaking only the time-reversal symmetry $\mathcal{T}$ does not change the Hamiltonian for the twisted bilayer system when interlayer tunneling $M[\u]$ only involves its minimal set of Fourier components. However, lifting the constraints enforced by time-reversal symmetry $\mathcal{T}$ and the layer-exchange symmetry $S$ will enable additional $i(w_5 \tau^0+w_6 \mu^3\tau^0)$ terms in $M_0$, as well as the other corresponding terms in the $M_{\k}$'s that are related to $M_0$ by spatial symmetries. With these additional terms, the original zero-energy Dirac cones will be gapped out. Note that the discussion here in this paragraph is restricted to the twisted bilayer system where the interlayer tunneling $M[\u]$ only includes its minimal set of the Fourier components. It would also be interesting to explore the effect of higher-momentum Fourier components (beyond the minimal set) in the interlayer tunneling.

In this paper, we have studied the case of the twisted-bilayer staggered-flux square lattice. The staggered-flux square lattice was initially introduced to characterize the mean-field band structure of spinons in an algebraic spin liquid. It would be interesting to consider bilayer systems with each layer describing the spinon band structure of other spin liquid candidates. In different spin liquids, the spinon bands has different symmetry properties. When it comes to a bilayer system with arbitrary elastic deformations in each layer, one can generalize the symmetry-based analysis to bootstrap the general form of interlayer tunneling for different spinon bands.

Our study paves the path towards understanding the twisted bilayer spin liquid with dynamical U(1) gauge field. For a spin liquid, it is a common wisdom that when there is only a small number of flavors of Dirac fermions in the spinon band structure, the monopoles of the dynamical gauge field have a tendency to drive the system into a confined phase. But with the extremely flat bands of the spinons derived in this work, the large density of states of the spinons at low energy might render the monopoles much less influential. Hence, deconfinement of spinon might happen over a large length scale if the physics discussed in this work is ever realized in real twisted magnetic materials. 

\bigskip

{\bf Acknowledgments:} Z.-X. L. is supported by the Simons
Collaborations on Ultra-Quantum Matter, grant 651440 (LB) from the Simons
Foundation. We thank Leon Balents for helpful discussions. C. X. is supported by NSF Grant No. DMR-1920434, and the Simons Foundation. Z.-X. L. is grateful to Michael Hermele and Jason Alicea for explaining their earlier works. C.-M. J. thanks Biao Lian for helpful discussion on band connectivity.

\bibliography{draft.bib}

\begin{thebibliography}{66}%
\makeatletter
\providecommand \@ifxundefined [1]{%
 \@ifx{#1\undefined}
}%
\providecommand \@ifnum [1]{%
 \ifnum #1\expandafter \@firstoftwo
 \else \expandafter \@secondoftwo
 \fi
}%
\providecommand \@ifx [1]{%
 \ifx #1\expandafter \@firstoftwo
 \else \expandafter \@secondoftwo
 \fi
}%
\providecommand \natexlab [1]{#1}%
\providecommand \enquote  [1]{``#1''}%
\providecommand \bibnamefont  [1]{#1}%
\providecommand \bibfnamefont [1]{#1}%
\providecommand \citenamefont [1]{#1}%
\providecommand \href@noop [0]{\@secondoftwo}%
\providecommand \href [0]{\begingroup \@sanitize@url \@href}%
\providecommand \@href[1]{\@@startlink{#1}\@@href}%
\providecommand \@@href[1]{\endgroup#1\@@endlink}%
\providecommand \@sanitize@url [0]{\catcode `\\12\catcode `\$12\catcode
  `\&12\catcode `\#12\catcode `\^12\catcode `\_12\catcode `\%12\relax}%
\providecommand \@@startlink[1]{}%
\providecommand \@@endlink[0]{}%
\providecommand \url  [0]{\begingroup\@sanitize@url \@url }%
\providecommand \@url [1]{\endgroup\@href {#1}{\urlprefix }}%
\providecommand \urlprefix  [0]{URL }%
\providecommand \Eprint [0]{\href }%
\providecommand \doibase [0]{http://dx.doi.org/}%
\providecommand \selectlanguage [0]{\@gobble}%
\providecommand \bibinfo  [0]{\@secondoftwo}%
\providecommand \bibfield  [0]{\@secondoftwo}%
\providecommand \translation [1]{[#1]}%
\providecommand \BibitemOpen [0]{}%
\providecommand \bibitemStop [0]{}%
\providecommand \bibitemNoStop [0]{.\EOS\space}%
\providecommand \EOS [0]{\spacefactor3000\relax}%
\providecommand \BibitemShut  [1]{\csname bibitem#1\endcsname}%
\let\auto@bib@innerbib\@empty
\bibitem [{\citenamefont {Cao}\ \emph {et~al.}(2018{\natexlab{a}})\citenamefont
  {Cao}, \citenamefont {Fatemi}, \citenamefont {Demir}, \citenamefont {Fang},
  \citenamefont {Tomarken}, \citenamefont {Luo}, \citenamefont
  {Sanchez-Yamagishi}, \citenamefont {Watanabe}, \citenamefont {Taniguchi},
  \citenamefont {Kaxiras}, \citenamefont {Ashoori},\ and\ \citenamefont
  {Jarillo-Herrero}}]{cao2018correlated}%
  \BibitemOpen
  \bibfield  {author} {\bibinfo {author} {\bibfnamefont {Y.}~\bibnamefont
  {Cao}}, \bibinfo {author} {\bibfnamefont {V.}~\bibnamefont {Fatemi}},
  \bibinfo {author} {\bibfnamefont {A.}~\bibnamefont {Demir}}, \bibinfo
  {author} {\bibfnamefont {S.}~\bibnamefont {Fang}}, \bibinfo {author}
  {\bibfnamefont {S.~L.}\ \bibnamefont {Tomarken}}, \bibinfo {author}
  {\bibfnamefont {J.~Y.}\ \bibnamefont {Luo}}, \bibinfo {author} {\bibfnamefont
  {J.~D.}\ \bibnamefont {Sanchez-Yamagishi}}, \bibinfo {author} {\bibfnamefont
  {K.}~\bibnamefont {Watanabe}}, \bibinfo {author} {\bibfnamefont
  {T.}~\bibnamefont {Taniguchi}}, \bibinfo {author} {\bibfnamefont
  {E.}~\bibnamefont {Kaxiras}}, \bibinfo {author} {\bibfnamefont {R.~C.}\
  \bibnamefont {Ashoori}}, \ and\ \bibinfo {author} {\bibfnamefont
  {P.}~\bibnamefont {Jarillo-Herrero}},\ }\href
  {https://doi.org/10.1038/nature26154} {\bibfield  {journal} {\bibinfo
  {journal} {Nature}\ }\textbf {\bibinfo {volume} {556}},\ \bibinfo {pages}
  {80} (\bibinfo {year} {2018}{\natexlab{a}})}\BibitemShut {NoStop}%
\bibitem [{\citenamefont {Cao}\ \emph {et~al.}(2018{\natexlab{b}})\citenamefont
  {Cao}, \citenamefont {Fatemi}, \citenamefont {Fang}, \citenamefont
  {Watanabe}, \citenamefont {Taniguchi}, \citenamefont {Kaxiras},\ and\
  \citenamefont {Jarillo-Herrero}}]{cao2018unconventional}%
  \BibitemOpen
  \bibfield  {author} {\bibinfo {author} {\bibfnamefont {Y.}~\bibnamefont
  {Cao}}, \bibinfo {author} {\bibfnamefont {V.}~\bibnamefont {Fatemi}},
  \bibinfo {author} {\bibfnamefont {S.}~\bibnamefont {Fang}}, \bibinfo {author}
  {\bibfnamefont {K.}~\bibnamefont {Watanabe}}, \bibinfo {author}
  {\bibfnamefont {T.}~\bibnamefont {Taniguchi}}, \bibinfo {author}
  {\bibfnamefont {E.}~\bibnamefont {Kaxiras}}, \ and\ \bibinfo {author}
  {\bibfnamefont {P.}~\bibnamefont {Jarillo-Herrero}},\ }\href
  {https://doi.org/10.1038/nature26160} {\bibfield  {journal} {\bibinfo
  {journal} {Nature}\ }\textbf {\bibinfo {volume} {556}},\ \bibinfo {pages}
  {43} (\bibinfo {year} {2018}{\natexlab{b}})}\BibitemShut {NoStop}%
\bibitem [{\citenamefont {Yankowitz}\ \emph {et~al.}(2019)\citenamefont
  {Yankowitz}, \citenamefont {Chen}, \citenamefont {Polshyn}, \citenamefont
  {Zhang}, \citenamefont {Watanabe}, \citenamefont {Taniguchi}, \citenamefont
  {Graf}, \citenamefont {Young},\ and\ \citenamefont
  {Dean}}]{yankowitz2019tuning}%
  \BibitemOpen
  \bibfield  {author} {\bibinfo {author} {\bibfnamefont {M.}~\bibnamefont
  {Yankowitz}}, \bibinfo {author} {\bibfnamefont {S.}~\bibnamefont {Chen}},
  \bibinfo {author} {\bibfnamefont {H.}~\bibnamefont {Polshyn}}, \bibinfo
  {author} {\bibfnamefont {Y.}~\bibnamefont {Zhang}}, \bibinfo {author}
  {\bibfnamefont {K.}~\bibnamefont {Watanabe}}, \bibinfo {author}
  {\bibfnamefont {T.}~\bibnamefont {Taniguchi}}, \bibinfo {author}
  {\bibfnamefont {D.}~\bibnamefont {Graf}}, \bibinfo {author} {\bibfnamefont
  {A.~F.}\ \bibnamefont {Young}}, \ and\ \bibinfo {author} {\bibfnamefont
  {C.~R.}\ \bibnamefont {Dean}},\ }\href
  {https://doi.org/10.1126/science.aav1910} {\bibfield  {journal} {\bibinfo
  {journal} {Science}\ }\textbf {\bibinfo {volume} {363}},\ \bibinfo {pages}
  {1059} (\bibinfo {year} {2019})}\BibitemShut {NoStop}%
\bibitem [{\citenamefont {Sharpe}\ \emph {et~al.}(2019)\citenamefont {Sharpe},
  \citenamefont {Fox}, \citenamefont {Barnard}, \citenamefont {Finney},
  \citenamefont {Watanabe}, \citenamefont {Taniguchi}, \citenamefont
  {Kastner},\ and\ \citenamefont {Goldhaber-Gordon}}]{sharpe2019emergent}%
  \BibitemOpen
  \bibfield  {author} {\bibinfo {author} {\bibfnamefont {A.~L.}\ \bibnamefont
  {Sharpe}}, \bibinfo {author} {\bibfnamefont {E.~J.}\ \bibnamefont {Fox}},
  \bibinfo {author} {\bibfnamefont {A.~W.}\ \bibnamefont {Barnard}}, \bibinfo
  {author} {\bibfnamefont {J.}~\bibnamefont {Finney}}, \bibinfo {author}
  {\bibfnamefont {K.}~\bibnamefont {Watanabe}}, \bibinfo {author}
  {\bibfnamefont {T.}~\bibnamefont {Taniguchi}}, \bibinfo {author}
  {\bibfnamefont {M.~A.}\ \bibnamefont {Kastner}}, \ and\ \bibinfo {author}
  {\bibfnamefont {D.}~\bibnamefont {Goldhaber-Gordon}},\ }\href
  {https://doi.org/10.1126/science.aaw3780} {\bibfield  {journal} {\bibinfo
  {journal} {Science}\ }\textbf {\bibinfo {volume} {365}},\ \bibinfo {pages}
  {605} (\bibinfo {year} {2019})}\BibitemShut {NoStop}%
\bibitem [{\citenamefont {Lu}\ \emph {et~al.}(2019)\citenamefont {Lu},
  \citenamefont {Stepanov}, \citenamefont {Yang}, \citenamefont {Xie},
  \citenamefont {Aamir}, \citenamefont {Das}, \citenamefont {Urgell},
  \citenamefont {Watanabe}, \citenamefont {Taniguchi}, \citenamefont {Zhang},
  \citenamefont {Bachtold}, \citenamefont {MacDonald},\ and\ \citenamefont
  {Efetov}}]{lu2019superconductors}%
  \BibitemOpen
  \bibfield  {author} {\bibinfo {author} {\bibfnamefont {X.}~\bibnamefont
  {Lu}}, \bibinfo {author} {\bibfnamefont {P.}~\bibnamefont {Stepanov}},
  \bibinfo {author} {\bibfnamefont {W.}~\bibnamefont {Yang}}, \bibinfo {author}
  {\bibfnamefont {M.}~\bibnamefont {Xie}}, \bibinfo {author} {\bibfnamefont
  {M.~A.}\ \bibnamefont {Aamir}}, \bibinfo {author} {\bibfnamefont
  {I.}~\bibnamefont {Das}}, \bibinfo {author} {\bibfnamefont {C.}~\bibnamefont
  {Urgell}}, \bibinfo {author} {\bibfnamefont {K.}~\bibnamefont {Watanabe}},
  \bibinfo {author} {\bibfnamefont {T.}~\bibnamefont {Taniguchi}}, \bibinfo
  {author} {\bibfnamefont {G.}~\bibnamefont {Zhang}}, \bibinfo {author}
  {\bibfnamefont {A.}~\bibnamefont {Bachtold}}, \bibinfo {author}
  {\bibfnamefont {A.~H.}\ \bibnamefont {MacDonald}}, \ and\ \bibinfo {author}
  {\bibfnamefont {D.~K.}\ \bibnamefont {Efetov}},\ }\href
  {https://doi.org/10.1038/s41586-019-1695-0} {\bibfield  {journal} {\bibinfo
  {journal} {Nature}\ }\textbf {\bibinfo {volume} {574}},\ \bibinfo {pages}
  {653} (\bibinfo {year} {2019})}\BibitemShut {NoStop}%
\bibitem [{\citenamefont {Serlin}\ \emph {et~al.}(2020)\citenamefont {Serlin},
  \citenamefont {Tschirhart}, \citenamefont {Polshyn}, \citenamefont {Zhang},
  \citenamefont {Zhu}, \citenamefont {Watanabe}, \citenamefont {Taniguchi},
  \citenamefont {Balents},\ and\ \citenamefont {Young}}]{serlin2020intrinsic}%
  \BibitemOpen
  \bibfield  {author} {\bibinfo {author} {\bibfnamefont {M.}~\bibnamefont
  {Serlin}}, \bibinfo {author} {\bibfnamefont {C.~L.}\ \bibnamefont
  {Tschirhart}}, \bibinfo {author} {\bibfnamefont {H.}~\bibnamefont {Polshyn}},
  \bibinfo {author} {\bibfnamefont {Y.}~\bibnamefont {Zhang}}, \bibinfo
  {author} {\bibfnamefont {J.}~\bibnamefont {Zhu}}, \bibinfo {author}
  {\bibfnamefont {K.}~\bibnamefont {Watanabe}}, \bibinfo {author}
  {\bibfnamefont {T.}~\bibnamefont {Taniguchi}}, \bibinfo {author}
  {\bibfnamefont {L.}~\bibnamefont {Balents}}, \ and\ \bibinfo {author}
  {\bibfnamefont {A.~F.}\ \bibnamefont {Young}},\ }\href
  {https://doi.org/10.1126/science.aay5533} {\bibfield  {journal} {\bibinfo
  {journal} {Science}\ }\textbf {\bibinfo {volume} {367}},\ \bibinfo {pages}
  {900} (\bibinfo {year} {2020})}\BibitemShut {NoStop}%
\bibitem [{\citenamefont {Kerelsky}\ \emph {et~al.}(2019)\citenamefont
  {Kerelsky}, \citenamefont {McGilly}, \citenamefont {Kennes}, \citenamefont
  {Xian}, \citenamefont {Yankowitz}, \citenamefont {Chen}, \citenamefont
  {Watanabe}, \citenamefont {Taniguchi}, \citenamefont {Hone}, \citenamefont
  {Dean}, \citenamefont {Rubio},\ and\ \citenamefont
  {Pasupathy}}]{kerelsky2019maximized}%
  \BibitemOpen
  \bibfield  {author} {\bibinfo {author} {\bibfnamefont {A.}~\bibnamefont
  {Kerelsky}}, \bibinfo {author} {\bibfnamefont {L.~J.}\ \bibnamefont
  {McGilly}}, \bibinfo {author} {\bibfnamefont {D.~M.}\ \bibnamefont {Kennes}},
  \bibinfo {author} {\bibfnamefont {L.}~\bibnamefont {Xian}}, \bibinfo {author}
  {\bibfnamefont {M.}~\bibnamefont {Yankowitz}}, \bibinfo {author}
  {\bibfnamefont {S.}~\bibnamefont {Chen}}, \bibinfo {author} {\bibfnamefont
  {K.}~\bibnamefont {Watanabe}}, \bibinfo {author} {\bibfnamefont
  {T.}~\bibnamefont {Taniguchi}}, \bibinfo {author} {\bibfnamefont
  {J.}~\bibnamefont {Hone}}, \bibinfo {author} {\bibfnamefont {C.}~\bibnamefont
  {Dean}}, \bibinfo {author} {\bibfnamefont {A.}~\bibnamefont {Rubio}}, \ and\
  \bibinfo {author} {\bibfnamefont {A.~N.}\ \bibnamefont {Pasupathy}},\ }\href
  {https://www.nature.com/articles/s41586-019-1431-9} {\bibfield  {journal}
  {\bibinfo  {journal} {Nature}\ }\textbf {\bibinfo {volume} {572}},\ \bibinfo
  {pages} {95} (\bibinfo {year} {2019})}\BibitemShut {NoStop}%
\bibitem [{\citenamefont {Xie}\ \emph {et~al.}(2019)\citenamefont {Xie},
  \citenamefont {Lian}, \citenamefont {J{\"a}ck}, \citenamefont {Liu},
  \citenamefont {Chiu}, \citenamefont {Watanabe}, \citenamefont {Taniguchi},
  \citenamefont {Bernevig},\ and\ \citenamefont
  {Yazdani}}]{xie2019spectroscopic}%
  \BibitemOpen
  \bibfield  {author} {\bibinfo {author} {\bibfnamefont {Y.}~\bibnamefont
  {Xie}}, \bibinfo {author} {\bibfnamefont {B.}~\bibnamefont {Lian}}, \bibinfo
  {author} {\bibfnamefont {B.}~\bibnamefont {J{\"a}ck}}, \bibinfo {author}
  {\bibfnamefont {X.}~\bibnamefont {Liu}}, \bibinfo {author} {\bibfnamefont
  {C.-L.}\ \bibnamefont {Chiu}}, \bibinfo {author} {\bibfnamefont
  {K.}~\bibnamefont {Watanabe}}, \bibinfo {author} {\bibfnamefont
  {T.}~\bibnamefont {Taniguchi}}, \bibinfo {author} {\bibfnamefont {B.~A.}\
  \bibnamefont {Bernevig}}, \ and\ \bibinfo {author} {\bibfnamefont
  {A.}~\bibnamefont {Yazdani}},\ }\href
  {https://doi.org/10.1038/s41586-019-1422-x} {\bibfield  {journal} {\bibinfo
  {journal} {Nature}\ }\textbf {\bibinfo {volume} {572}},\ \bibinfo {pages}
  {101} (\bibinfo {year} {2019})}\BibitemShut {NoStop}%
\bibitem [{\citenamefont {Jiang}\ \emph {et~al.}(2019)\citenamefont {Jiang},
  \citenamefont {Lai}, \citenamefont {Watanabe}, \citenamefont {Taniguchi},
  \citenamefont {Haule}, \citenamefont {Mao},\ and\ \citenamefont
  {Andrei}}]{jiang2019charge}%
  \BibitemOpen
  \bibfield  {author} {\bibinfo {author} {\bibfnamefont {Y.}~\bibnamefont
  {Jiang}}, \bibinfo {author} {\bibfnamefont {X.}~\bibnamefont {Lai}}, \bibinfo
  {author} {\bibfnamefont {K.}~\bibnamefont {Watanabe}}, \bibinfo {author}
  {\bibfnamefont {T.}~\bibnamefont {Taniguchi}}, \bibinfo {author}
  {\bibfnamefont {K.}~\bibnamefont {Haule}}, \bibinfo {author} {\bibfnamefont
  {J.}~\bibnamefont {Mao}}, \ and\ \bibinfo {author} {\bibfnamefont {E.~Y.}\
  \bibnamefont {Andrei}},\ }\href {https://doi.org/10.1038/s41586-019-1460-4}
  {\bibfield  {journal} {\bibinfo  {journal} {Nature}\ }\textbf {\bibinfo
  {volume} {573}},\ \bibinfo {pages} {91} (\bibinfo {year} {2019})}\BibitemShut
  {NoStop}%
\bibitem [{\citenamefont {Carr}\ \emph {et~al.}(2017)\citenamefont {Carr},
  \citenamefont {Massatt}, \citenamefont {Fang}, \citenamefont {Cazeaux},
  \citenamefont {Luskin},\ and\ \citenamefont {Kaxiras}}]{carr2017twistronics}%
  \BibitemOpen
  \bibfield  {author} {\bibinfo {author} {\bibfnamefont {S.}~\bibnamefont
  {Carr}}, \bibinfo {author} {\bibfnamefont {D.}~\bibnamefont {Massatt}},
  \bibinfo {author} {\bibfnamefont {S.}~\bibnamefont {Fang}}, \bibinfo {author}
  {\bibfnamefont {P.}~\bibnamefont {Cazeaux}}, \bibinfo {author} {\bibfnamefont
  {M.}~\bibnamefont {Luskin}}, \ and\ \bibinfo {author} {\bibfnamefont
  {E.}~\bibnamefont {Kaxiras}},\ }\href
  {https://doi.org/10.1103/PhysRevB.95.075420} {\bibfield  {journal} {\bibinfo
  {journal} {Physical Review B}\ }\textbf {\bibinfo {volume} {95}},\ \bibinfo
  {pages} {075420} (\bibinfo {year} {2017})}\BibitemShut {NoStop}%
\bibitem [{\citenamefont {Tran}\ \emph {et~al.}(2019)\citenamefont {Tran},
  \citenamefont {Moody}, \citenamefont {Wu}, \citenamefont {Lu}, \citenamefont
  {Choi}, \citenamefont {Kim}, \citenamefont {Rai}, \citenamefont {Sanchez},
  \citenamefont {Quan}, \citenamefont {Singh}, \citenamefont {Embley},
  \citenamefont {Zepeda}, \citenamefont {Campbell}, \citenamefont {Autry},
  \citenamefont {Taniguchi}, \citenamefont {Watanabe}, \citenamefont {Lu},
  \citenamefont {Banerjee}, \citenamefont {Silverman}, \citenamefont {Kim},
  \citenamefont {Tutuc}, \citenamefont {Yang}, \citenamefont {MacDonald},\ and\
  \citenamefont {Li}}]{tran2019evidence}%
  \BibitemOpen
  \bibfield  {author} {\bibinfo {author} {\bibfnamefont {K.}~\bibnamefont
  {Tran}}, \bibinfo {author} {\bibfnamefont {G.}~\bibnamefont {Moody}},
  \bibinfo {author} {\bibfnamefont {F.}~\bibnamefont {Wu}}, \bibinfo {author}
  {\bibfnamefont {X.}~\bibnamefont {Lu}}, \bibinfo {author} {\bibfnamefont
  {J.}~\bibnamefont {Choi}}, \bibinfo {author} {\bibfnamefont {K.}~\bibnamefont
  {Kim}}, \bibinfo {author} {\bibfnamefont {A.}~\bibnamefont {Rai}}, \bibinfo
  {author} {\bibfnamefont {D.~A.}\ \bibnamefont {Sanchez}}, \bibinfo {author}
  {\bibfnamefont {J.}~\bibnamefont {Quan}}, \bibinfo {author} {\bibfnamefont
  {A.}~\bibnamefont {Singh}}, \bibinfo {author} {\bibfnamefont
  {J.}~\bibnamefont {Embley}}, \bibinfo {author} {\bibfnamefont
  {A.}~\bibnamefont {Zepeda}}, \bibinfo {author} {\bibfnamefont
  {M.}~\bibnamefont {Campbell}}, \bibinfo {author} {\bibfnamefont
  {T.}~\bibnamefont {Autry}}, \bibinfo {author} {\bibfnamefont
  {T.}~\bibnamefont {Taniguchi}}, \bibinfo {author} {\bibfnamefont
  {K.}~\bibnamefont {Watanabe}}, \bibinfo {author} {\bibfnamefont
  {N.}~\bibnamefont {Lu}}, \bibinfo {author} {\bibfnamefont {S.~K.}\
  \bibnamefont {Banerjee}}, \bibinfo {author} {\bibfnamefont {K.~L.}\
  \bibnamefont {Silverman}}, \bibinfo {author} {\bibfnamefont {S.}~\bibnamefont
  {Kim}}, \bibinfo {author} {\bibfnamefont {E.}~\bibnamefont {Tutuc}}, \bibinfo
  {author} {\bibfnamefont {L.}~\bibnamefont {Yang}}, \bibinfo {author}
  {\bibfnamefont {A.~H.}\ \bibnamefont {MacDonald}}, \ and\ \bibinfo {author}
  {\bibfnamefont {X.}~\bibnamefont {Li}},\ }\href
  {https://doi.org/10.1038/s41586-019-0975-z} {\bibfield  {journal} {\bibinfo
  {journal} {Nature}\ }\textbf {\bibinfo {volume} {567}},\ \bibinfo {pages}
  {71} (\bibinfo {year} {2019})}\BibitemShut {NoStop}%
\bibitem [{\citenamefont {Jin}\ \emph {et~al.}(2019)\citenamefont {Jin},
  \citenamefont {Regan}, \citenamefont {Yan}, \citenamefont {Utama},
  \citenamefont {Wang}, \citenamefont {Zhao}, \citenamefont {Qin},
  \citenamefont {Yang}, \citenamefont {Zheng}, \citenamefont {Shi},
  \citenamefont {Watanabe}, \citenamefont {Taniguchi}, \citenamefont {Tongay},
  \citenamefont {Zettl},\ and\ \citenamefont {Wang}}]{jin2019observation}%
  \BibitemOpen
  \bibfield  {author} {\bibinfo {author} {\bibfnamefont {C.}~\bibnamefont
  {Jin}}, \bibinfo {author} {\bibfnamefont {E.~C.}\ \bibnamefont {Regan}},
  \bibinfo {author} {\bibfnamefont {A.}~\bibnamefont {Yan}}, \bibinfo {author}
  {\bibfnamefont {M.~I.~B.}\ \bibnamefont {Utama}}, \bibinfo {author}
  {\bibfnamefont {D.}~\bibnamefont {Wang}}, \bibinfo {author} {\bibfnamefont
  {S.}~\bibnamefont {Zhao}}, \bibinfo {author} {\bibfnamefont {Y.}~\bibnamefont
  {Qin}}, \bibinfo {author} {\bibfnamefont {S.}~\bibnamefont {Yang}}, \bibinfo
  {author} {\bibfnamefont {Z.}~\bibnamefont {Zheng}}, \bibinfo {author}
  {\bibfnamefont {S.}~\bibnamefont {Shi}}, \bibinfo {author} {\bibfnamefont
  {K.}~\bibnamefont {Watanabe}}, \bibinfo {author} {\bibfnamefont
  {T.}~\bibnamefont {Taniguchi}}, \bibinfo {author} {\bibfnamefont
  {S.}~\bibnamefont {Tongay}}, \bibinfo {author} {\bibfnamefont
  {A.}~\bibnamefont {Zettl}}, \ and\ \bibinfo {author} {\bibfnamefont
  {F.}~\bibnamefont {Wang}},\ }\href
  {https://doi.org/10.1038/s41586-019-0976-y} {\bibfield  {journal} {\bibinfo
  {journal} {Nature}\ }\textbf {\bibinfo {volume} {567}},\ \bibinfo {pages}
  {76} (\bibinfo {year} {2019})}\BibitemShut {NoStop}%
\bibitem [{\citenamefont {Seyler}\ \emph {et~al.}(2019)\citenamefont {Seyler},
  \citenamefont {Rivera}, \citenamefont {Yu}, \citenamefont {Wilson},
  \citenamefont {Ray}, \citenamefont {Mandrus}, \citenamefont {Yan},
  \citenamefont {Yao},\ and\ \citenamefont {Xu}}]{seyler2019signatures}%
  \BibitemOpen
  \bibfield  {author} {\bibinfo {author} {\bibfnamefont {K.~L.}\ \bibnamefont
  {Seyler}}, \bibinfo {author} {\bibfnamefont {P.}~\bibnamefont {Rivera}},
  \bibinfo {author} {\bibfnamefont {H.}~\bibnamefont {Yu}}, \bibinfo {author}
  {\bibfnamefont {N.~P.}\ \bibnamefont {Wilson}}, \bibinfo {author}
  {\bibfnamefont {E.~L.}\ \bibnamefont {Ray}}, \bibinfo {author} {\bibfnamefont
  {D.~G.}\ \bibnamefont {Mandrus}}, \bibinfo {author} {\bibfnamefont
  {J.}~\bibnamefont {Yan}}, \bibinfo {author} {\bibfnamefont {W.}~\bibnamefont
  {Yao}}, \ and\ \bibinfo {author} {\bibfnamefont {X.}~\bibnamefont {Xu}},\
  }\href {https://doi.org/10.1038/s41586-019-0957-1} {\bibfield  {journal}
  {\bibinfo  {journal} {Nature}\ }\textbf {\bibinfo {volume} {567}},\ \bibinfo
  {pages} {66} (\bibinfo {year} {2019})}\BibitemShut {NoStop}%
\bibitem [{\citenamefont {Alexeev}\ \emph {et~al.}(2019)\citenamefont
  {Alexeev}, \citenamefont {Ruiz-Tijerina}, \citenamefont {Danovich},
  \citenamefont {Hamer}, \citenamefont {Terry}, \citenamefont {Nayak},
  \citenamefont {Ahn}, \citenamefont {Pak}, \citenamefont {Lee}, \citenamefont
  {Sohn}, \citenamefont {Molas}, \citenamefont {Koperski}, \citenamefont
  {Watanabe}, \citenamefont {Taniguchi}, \citenamefont {Novoselov},
  \citenamefont {Gorbachev}, \citenamefont {Shin}, \citenamefont {Fal'ko},\
  and\ \citenamefont {Tartakovskii}}]{alexeev2019resonantly}%
  \BibitemOpen
  \bibfield  {author} {\bibinfo {author} {\bibfnamefont {E.~M.}\ \bibnamefont
  {Alexeev}}, \bibinfo {author} {\bibfnamefont {D.~A.}\ \bibnamefont
  {Ruiz-Tijerina}}, \bibinfo {author} {\bibfnamefont {M.}~\bibnamefont
  {Danovich}}, \bibinfo {author} {\bibfnamefont {M.~J.}\ \bibnamefont {Hamer}},
  \bibinfo {author} {\bibfnamefont {D.~J.}\ \bibnamefont {Terry}}, \bibinfo
  {author} {\bibfnamefont {P.~K.}\ \bibnamefont {Nayak}}, \bibinfo {author}
  {\bibfnamefont {S.}~\bibnamefont {Ahn}}, \bibinfo {author} {\bibfnamefont
  {S.}~\bibnamefont {Pak}}, \bibinfo {author} {\bibfnamefont {J.}~\bibnamefont
  {Lee}}, \bibinfo {author} {\bibfnamefont {J.~I.}\ \bibnamefont {Sohn}},
  \bibinfo {author} {\bibfnamefont {M.~R.}\ \bibnamefont {Molas}}, \bibinfo
  {author} {\bibfnamefont {M.}~\bibnamefont {Koperski}}, \bibinfo {author}
  {\bibfnamefont {K.}~\bibnamefont {Watanabe}}, \bibinfo {author}
  {\bibfnamefont {T.}~\bibnamefont {Taniguchi}}, \bibinfo {author}
  {\bibfnamefont {K.~S.}\ \bibnamefont {Novoselov}}, \bibinfo {author}
  {\bibfnamefont {R.~V.}\ \bibnamefont {Gorbachev}}, \bibinfo {author}
  {\bibfnamefont {H.~S.}\ \bibnamefont {Shin}}, \bibinfo {author}
  {\bibfnamefont {V.~I.}\ \bibnamefont {Fal'ko}}, \ and\ \bibinfo {author}
  {\bibfnamefont {A.~I.}\ \bibnamefont {Tartakovskii}},\ }\href
  {https://doi.org/10.1038/s41586-019-0986-9} {\bibfield  {journal} {\bibinfo
  {journal} {Nature}\ }\textbf {\bibinfo {volume} {567}},\ \bibinfo {pages}
  {81} (\bibinfo {year} {2019})}\BibitemShut {NoStop}%
\bibitem [{\citenamefont {Tang}\ \emph {et~al.}(2020)\citenamefont {Tang},
  \citenamefont {Li}, \citenamefont {Li}, \citenamefont {Xu}, \citenamefont
  {Liu}, \citenamefont {Barmak}, \citenamefont {Watanabe}, \citenamefont
  {Taniguchi}, \citenamefont {MacDonald}, \citenamefont {Shan},\ and\
  \citenamefont {Mak}}]{tang2020simulation}%
  \BibitemOpen
  \bibfield  {author} {\bibinfo {author} {\bibfnamefont {Y.}~\bibnamefont
  {Tang}}, \bibinfo {author} {\bibfnamefont {L.}~\bibnamefont {Li}}, \bibinfo
  {author} {\bibfnamefont {T.}~\bibnamefont {Li}}, \bibinfo {author}
  {\bibfnamefont {Y.}~\bibnamefont {Xu}}, \bibinfo {author} {\bibfnamefont
  {S.}~\bibnamefont {Liu}}, \bibinfo {author} {\bibfnamefont {K.}~\bibnamefont
  {Barmak}}, \bibinfo {author} {\bibfnamefont {K.}~\bibnamefont {Watanabe}},
  \bibinfo {author} {\bibfnamefont {T.}~\bibnamefont {Taniguchi}}, \bibinfo
  {author} {\bibfnamefont {A.~H.}\ \bibnamefont {MacDonald}}, \bibinfo {author}
  {\bibfnamefont {J.}~\bibnamefont {Shan}}, \ and\ \bibinfo {author}
  {\bibfnamefont {K.~F.}\ \bibnamefont {Mak}},\ }\href
  {https://doi.org/10.1038/s41586-020-2085-3} {\bibfield  {journal} {\bibinfo
  {journal} {Nature}\ }\textbf {\bibinfo {volume} {579}},\ \bibinfo {pages}
  {353} (\bibinfo {year} {2020})}\BibitemShut {NoStop}%
\bibitem [{\citenamefont {Regan}\ \emph {et~al.}(2020)\citenamefont {Regan},
  \citenamefont {Wang}, \citenamefont {Jin}, \citenamefont {Utama},
  \citenamefont {Gao}, \citenamefont {Wei}, \citenamefont {Zhao}, \citenamefont
  {Zhao}, \citenamefont {Zhang}, \citenamefont {Yumigeta}, \citenamefont
  {Blei}, \citenamefont {Carlstr\"om}, \citenamefont {Watanabe}, \citenamefont
  {Taniguchi}, \citenamefont {Tongay}, \citenamefont {Michael}, \citenamefont
  {Zettl},\ and\ \citenamefont {Wang}}]{regan2020mott}%
  \BibitemOpen
  \bibfield  {author} {\bibinfo {author} {\bibfnamefont {E.~C.}\ \bibnamefont
  {Regan}}, \bibinfo {author} {\bibfnamefont {D.}~\bibnamefont {Wang}},
  \bibinfo {author} {\bibfnamefont {C.}~\bibnamefont {Jin}}, \bibinfo {author}
  {\bibfnamefont {M.~I.~B.}\ \bibnamefont {Utama}}, \bibinfo {author}
  {\bibfnamefont {B.}~\bibnamefont {Gao}}, \bibinfo {author} {\bibfnamefont
  {X.}~\bibnamefont {Wei}}, \bibinfo {author} {\bibfnamefont {S.}~\bibnamefont
  {Zhao}}, \bibinfo {author} {\bibfnamefont {W.}~\bibnamefont {Zhao}}, \bibinfo
  {author} {\bibfnamefont {Z.}~\bibnamefont {Zhang}}, \bibinfo {author}
  {\bibfnamefont {K.}~\bibnamefont {Yumigeta}}, \bibinfo {author}
  {\bibfnamefont {M.}~\bibnamefont {Blei}}, \bibinfo {author} {\bibfnamefont
  {J.~D.}\ \bibnamefont {Carlstr\"om}}, \bibinfo {author} {\bibfnamefont
  {K.}~\bibnamefont {Watanabe}}, \bibinfo {author} {\bibfnamefont
  {T.}~\bibnamefont {Taniguchi}}, \bibinfo {author} {\bibfnamefont
  {S.}~\bibnamefont {Tongay}}, \bibinfo {author} {\bibfnamefont
  {C.}~\bibnamefont {Michael}}, \bibinfo {author} {\bibfnamefont
  {A.}~\bibnamefont {Zettl}}, \ and\ \bibinfo {author} {\bibfnamefont
  {F.}~\bibnamefont {Wang}},\ }\href
  {https://doi.org/10.1038/s41586-020-2092-4} {\bibfield  {journal} {\bibinfo
  {journal} {Nature}\ }\textbf {\bibinfo {volume} {579}},\ \bibinfo {pages}
  {359} (\bibinfo {year} {2020})}\BibitemShut {NoStop}%
\bibitem [{\citenamefont {Shimazaki}\ \emph {et~al.}(2020)\citenamefont
  {Shimazaki}, \citenamefont {Schwartz}, \citenamefont {Watanabe},
  \citenamefont {Taniguchi}, \citenamefont {Kroner},\ and\ \citenamefont
  {Imamoglu}}]{shimazaki2020strongly}%
  \BibitemOpen
  \bibfield  {author} {\bibinfo {author} {\bibfnamefont {Y.}~\bibnamefont
  {Shimazaki}}, \bibinfo {author} {\bibfnamefont {I.}~\bibnamefont {Schwartz}},
  \bibinfo {author} {\bibfnamefont {K.}~\bibnamefont {Watanabe}}, \bibinfo
  {author} {\bibfnamefont {T.}~\bibnamefont {Taniguchi}}, \bibinfo {author}
  {\bibfnamefont {M.}~\bibnamefont {Kroner}}, \ and\ \bibinfo {author}
  {\bibfnamefont {A.}~\bibnamefont {Imamoglu}},\ }\href
  {https://doi.org/10.1038/s41586-020-2191-2} {\bibfield  {journal} {\bibinfo
  {journal} {Nature}\ }\textbf {\bibinfo {volume} {580}},\ \bibinfo {pages}
  {472} (\bibinfo {year} {2020})}\BibitemShut {NoStop}%
\bibitem [{\citenamefont {Wang}\ \emph
  {et~al.}(2020{\natexlab{a}})\citenamefont {Wang}, \citenamefont {Shih},
  \citenamefont {Ghiotto}, \citenamefont {Xian}, \citenamefont {Rhodes},
  \citenamefont {Tan}, \citenamefont {Claassen}, \citenamefont {Kennes},
  \citenamefont {Bai}, \citenamefont {Kim}, \citenamefont {Watanabe},
  \citenamefont {Taniguchi}, \citenamefont {Zhu}, \citenamefont {Hone},
  \citenamefont {Rubio}, \citenamefont {Pasupathy},\ and\ \citenamefont
  {Dean}}]{wang2020correlated}%
  \BibitemOpen
  \bibfield  {author} {\bibinfo {author} {\bibfnamefont {L.}~\bibnamefont
  {Wang}}, \bibinfo {author} {\bibfnamefont {E.-M.}\ \bibnamefont {Shih}},
  \bibinfo {author} {\bibfnamefont {A.}~\bibnamefont {Ghiotto}}, \bibinfo
  {author} {\bibfnamefont {L.}~\bibnamefont {Xian}}, \bibinfo {author}
  {\bibfnamefont {D.~A.}\ \bibnamefont {Rhodes}}, \bibinfo {author}
  {\bibfnamefont {C.}~\bibnamefont {Tan}}, \bibinfo {author} {\bibfnamefont
  {M.}~\bibnamefont {Claassen}}, \bibinfo {author} {\bibfnamefont {D.~M.}\
  \bibnamefont {Kennes}}, \bibinfo {author} {\bibfnamefont {Y.}~\bibnamefont
  {Bai}}, \bibinfo {author} {\bibfnamefont {B.}~\bibnamefont {Kim}}, \bibinfo
  {author} {\bibfnamefont {K.}~\bibnamefont {Watanabe}}, \bibinfo {author}
  {\bibfnamefont {T.}~\bibnamefont {Taniguchi}}, \bibinfo {author}
  {\bibfnamefont {X.}~\bibnamefont {Zhu}}, \bibinfo {author} {\bibfnamefont
  {J.}~\bibnamefont {Hone}}, \bibinfo {author} {\bibfnamefont {A.}~\bibnamefont
  {Rubio}}, \bibinfo {author} {\bibfnamefont {A.~N.}\ \bibnamefont
  {Pasupathy}}, \ and\ \bibinfo {author} {\bibfnamefont {C.~R.}\ \bibnamefont
  {Dean}},\ }\href {\doibase 10.1038/s41563-020-0708-6} {\bibfield  {journal}
  {\bibinfo  {journal} {Nature Materials}\ }\textbf {\bibinfo {volume} {19}},\
  \bibinfo {pages} {861} (\bibinfo {year} {2020}{\natexlab{a}})}\BibitemShut
  {NoStop}%
\bibitem [{\citenamefont {Liu}\ \emph {et~al.}(2020)\citenamefont {Liu},
  \citenamefont {Hao}, \citenamefont {Khalaf}, \citenamefont {Lee},
  \citenamefont {Ronen}, \citenamefont {Yoo}, \citenamefont {Najafabadi},
  \citenamefont {Watanabe}, \citenamefont {Taniguchi}, \citenamefont
  {Vishwanath},\ and\ \citenamefont {Kim}}]{liu2020tunable}%
  \BibitemOpen
  \bibfield  {author} {\bibinfo {author} {\bibfnamefont {X.}~\bibnamefont
  {Liu}}, \bibinfo {author} {\bibfnamefont {Z.}~\bibnamefont {Hao}}, \bibinfo
  {author} {\bibfnamefont {E.}~\bibnamefont {Khalaf}}, \bibinfo {author}
  {\bibfnamefont {J.~Y.}\ \bibnamefont {Lee}}, \bibinfo {author} {\bibfnamefont
  {Y.}~\bibnamefont {Ronen}}, \bibinfo {author} {\bibfnamefont
  {H.}~\bibnamefont {Yoo}}, \bibinfo {author} {\bibfnamefont {D.~H.}\
  \bibnamefont {Najafabadi}}, \bibinfo {author} {\bibfnamefont
  {K.}~\bibnamefont {Watanabe}}, \bibinfo {author} {\bibfnamefont
  {T.}~\bibnamefont {Taniguchi}}, \bibinfo {author} {\bibfnamefont
  {A.}~\bibnamefont {Vishwanath}}, \ and\ \bibinfo {author} {\bibfnamefont
  {P.}~\bibnamefont {Kim}},\ }\href {https://doi.org/10.1038/s41586-020-2458-7}
  {\bibfield  {journal} {\bibinfo  {journal} {Nature}\ }\textbf {\bibinfo
  {volume} {583}},\ \bibinfo {pages} {221} (\bibinfo {year}
  {2020})}\BibitemShut {NoStop}%
\bibitem [{\citenamefont {Burg}\ \emph {et~al.}(2019)\citenamefont {Burg},
  \citenamefont {Zhu}, \citenamefont {Taniguchi}, \citenamefont {Watanabe},
  \citenamefont {MacDonald},\ and\ \citenamefont {Tutuc}}]{burg2019correlated}%
  \BibitemOpen
  \bibfield  {author} {\bibinfo {author} {\bibfnamefont {G.~W.}\ \bibnamefont
  {Burg}}, \bibinfo {author} {\bibfnamefont {J.}~\bibnamefont {Zhu}}, \bibinfo
  {author} {\bibfnamefont {T.}~\bibnamefont {Taniguchi}}, \bibinfo {author}
  {\bibfnamefont {K.}~\bibnamefont {Watanabe}}, \bibinfo {author}
  {\bibfnamefont {A.~H.}\ \bibnamefont {MacDonald}}, \ and\ \bibinfo {author}
  {\bibfnamefont {E.}~\bibnamefont {Tutuc}},\ }\href
  {https://doi.org/10.1103/PhysRevLett.123.197702} {\bibfield  {journal}
  {\bibinfo  {journal} {Physical review letters}\ }\textbf {\bibinfo {volume}
  {123}},\ \bibinfo {pages} {197702} (\bibinfo {year} {2019})}\BibitemShut
  {NoStop}%
\bibitem [{\citenamefont {Cao}\ \emph {et~al.}(2020)\citenamefont {Cao},
  \citenamefont {Rodan-Legrain}, \citenamefont {Rubies-Bigorda}, \citenamefont
  {Park}, \citenamefont {Watanabe}, \citenamefont {Taniguchi},\ and\
  \citenamefont {Jarillo-Herrero}}]{cao2020tunable}%
  \BibitemOpen
  \bibfield  {author} {\bibinfo {author} {\bibfnamefont {Y.}~\bibnamefont
  {Cao}}, \bibinfo {author} {\bibfnamefont {D.}~\bibnamefont {Rodan-Legrain}},
  \bibinfo {author} {\bibfnamefont {O.}~\bibnamefont {Rubies-Bigorda}},
  \bibinfo {author} {\bibfnamefont {J.~M.}\ \bibnamefont {Park}}, \bibinfo
  {author} {\bibfnamefont {K.}~\bibnamefont {Watanabe}}, \bibinfo {author}
  {\bibfnamefont {T.}~\bibnamefont {Taniguchi}}, \ and\ \bibinfo {author}
  {\bibfnamefont {P.}~\bibnamefont {Jarillo-Herrero}},\ }\href
  {https://doi.org/10.1038/s41586-020-2260-6} {\bibfield  {journal} {\bibinfo
  {journal} {Nature}\ }\textbf {\bibinfo {volume} {583}},\ \bibinfo {pages}
  {215} (\bibinfo {year} {2020})}\BibitemShut {NoStop}%
\bibitem [{\citenamefont {Shen}\ \emph {et~al.}(2020)\citenamefont {Shen},
  \citenamefont {Chu}, \citenamefont {Wu}, \citenamefont {Li}, \citenamefont
  {Wang}, \citenamefont {Zhao}, \citenamefont {Tang}, \citenamefont {Liu},
  \citenamefont {Tian}, \citenamefont {Watanabe}, , \citenamefont {Taniguchi},
  \citenamefont {Yang}, \citenamefont {Meng}, \citenamefont {Shi},
  \citenamefont {Yazyev},\ and\ \citenamefont {Zhang}}]{shen2020correlated}%
  \BibitemOpen
  \bibfield  {author} {\bibinfo {author} {\bibfnamefont {C.}~\bibnamefont
  {Shen}}, \bibinfo {author} {\bibfnamefont {Y.}~\bibnamefont {Chu}}, \bibinfo
  {author} {\bibfnamefont {Q.}~\bibnamefont {Wu}}, \bibinfo {author}
  {\bibfnamefont {N.}~\bibnamefont {Li}}, \bibinfo {author} {\bibfnamefont
  {S.}~\bibnamefont {Wang}}, \bibinfo {author} {\bibfnamefont {Y.}~\bibnamefont
  {Zhao}}, \bibinfo {author} {\bibfnamefont {J.}~\bibnamefont {Tang}}, \bibinfo
  {author} {\bibfnamefont {J.}~\bibnamefont {Liu}}, \bibinfo {author}
  {\bibfnamefont {J.}~\bibnamefont {Tian}}, \bibinfo {author} {\bibfnamefont
  {K.}~\bibnamefont {Watanabe}}, , \bibinfo {author} {\bibfnamefont
  {T.}~\bibnamefont {Taniguchi}}, \bibinfo {author} {\bibfnamefont
  {R.}~\bibnamefont {Yang}}, \bibinfo {author} {\bibfnamefont {Z.~Y.}\
  \bibnamefont {Meng}}, \bibinfo {author} {\bibfnamefont {D.}~\bibnamefont
  {Shi}}, \bibinfo {author} {\bibfnamefont {O.~V.}\ \bibnamefont {Yazyev}}, \
  and\ \bibinfo {author} {\bibfnamefont {G.}~\bibnamefont {Zhang}},\ }\href
  {https://doi.org/10.1038/s41567-020-0825-9} {\bibfield  {journal} {\bibinfo
  {journal} {Nature Physics}\ }\textbf {\bibinfo {volume} {16}},\ \bibinfo
  {pages} {520} (\bibinfo {year} {2020})}\BibitemShut {NoStop}%
\bibitem [{\citenamefont {Chen}\ \emph
  {et~al.}(2019{\natexlab{a}})\citenamefont {Chen}, \citenamefont {Jiang},
  \citenamefont {Wu}, \citenamefont {Lyu}, \citenamefont {Li}, \citenamefont
  {Chittari}, \citenamefont {Watanabe}, \citenamefont {Taniguchi},
  \citenamefont {Shi}, \citenamefont {Jung}, \citenamefont {Zhang},\ and\
  \citenamefont {Wang}}]{chen2019evidence}%
  \BibitemOpen
  \bibfield  {author} {\bibinfo {author} {\bibfnamefont {G.}~\bibnamefont
  {Chen}}, \bibinfo {author} {\bibfnamefont {L.}~\bibnamefont {Jiang}},
  \bibinfo {author} {\bibfnamefont {S.}~\bibnamefont {Wu}}, \bibinfo {author}
  {\bibfnamefont {B.}~\bibnamefont {Lyu}}, \bibinfo {author} {\bibfnamefont
  {H.}~\bibnamefont {Li}}, \bibinfo {author} {\bibfnamefont {B.~L.}\
  \bibnamefont {Chittari}}, \bibinfo {author} {\bibfnamefont {K.}~\bibnamefont
  {Watanabe}}, \bibinfo {author} {\bibfnamefont {T.}~\bibnamefont {Taniguchi}},
  \bibinfo {author} {\bibfnamefont {Z.}~\bibnamefont {Shi}}, \bibinfo {author}
  {\bibfnamefont {J.}~\bibnamefont {Jung}}, \bibinfo {author} {\bibfnamefont
  {Y.}~\bibnamefont {Zhang}}, \ and\ \bibinfo {author} {\bibfnamefont
  {F.}~\bibnamefont {Wang}},\ }\href
  {https://doi.org/10.1038/s41567-018-0387-2} {\bibfield  {journal} {\bibinfo
  {journal} {Nature Physics}\ }\textbf {\bibinfo {volume} {15}},\ \bibinfo
  {pages} {237} (\bibinfo {year} {2019}{\natexlab{a}})}\BibitemShut {NoStop}%
\bibitem [{\citenamefont {Chen}\ \emph
  {et~al.}(2019{\natexlab{b}})\citenamefont {Chen}, \citenamefont {Sharpe},
  \citenamefont {Gallagher}, \citenamefont {Rosen}, \citenamefont {Fox},
  \citenamefont {Jiang}, \citenamefont {Lyu}, \citenamefont {Li}, \citenamefont
  {Watanabe}, \citenamefont {Taniguchi}, \citenamefont {Jung}, \citenamefont
  {Shi}, \citenamefont {Goldhaber-Gordon}, \citenamefont {Zhang},\ and\
  \citenamefont {Wang}}]{chen2019signatures}%
  \BibitemOpen
  \bibfield  {author} {\bibinfo {author} {\bibfnamefont {G.}~\bibnamefont
  {Chen}}, \bibinfo {author} {\bibfnamefont {A.~L.}\ \bibnamefont {Sharpe}},
  \bibinfo {author} {\bibfnamefont {P.}~\bibnamefont {Gallagher}}, \bibinfo
  {author} {\bibfnamefont {I.~T.}\ \bibnamefont {Rosen}}, \bibinfo {author}
  {\bibfnamefont {E.~J.}\ \bibnamefont {Fox}}, \bibinfo {author} {\bibfnamefont
  {L.}~\bibnamefont {Jiang}}, \bibinfo {author} {\bibfnamefont
  {B.}~\bibnamefont {Lyu}}, \bibinfo {author} {\bibfnamefont {H.}~\bibnamefont
  {Li}}, \bibinfo {author} {\bibfnamefont {K.}~\bibnamefont {Watanabe}},
  \bibinfo {author} {\bibfnamefont {T.}~\bibnamefont {Taniguchi}}, \bibinfo
  {author} {\bibfnamefont {J.}~\bibnamefont {Jung}}, \bibinfo {author}
  {\bibfnamefont {Z.}~\bibnamefont {Shi}}, \bibinfo {author} {\bibfnamefont
  {D.}~\bibnamefont {Goldhaber-Gordon}}, \bibinfo {author} {\bibfnamefont
  {Y.}~\bibnamefont {Zhang}}, \ and\ \bibinfo {author} {\bibfnamefont
  {F.}~\bibnamefont {Wang}},\ }\href
  {https://doi.org/10.1038/s41586-019-1393-y} {\bibfield  {journal} {\bibinfo
  {journal} {Nature}\ }\textbf {\bibinfo {volume} {572}},\ \bibinfo {pages}
  {215} (\bibinfo {year} {2019}{\natexlab{b}})}\BibitemShut {NoStop}%
\bibitem [{\citenamefont {Chen}\ \emph {et~al.}(2020)\citenamefont {Chen},
  \citenamefont {Sharpe}, \citenamefont {Fox}, \citenamefont {Zhang},
  \citenamefont {Wang}, \citenamefont {Jiang}, \citenamefont {Lyu},
  \citenamefont {Li}, \citenamefont {Watanabe}, \citenamefont {Taniguchi},
  \citenamefont {Shi}, \citenamefont {Senthil}, \citenamefont
  {Goldhaber-Gordon}, \citenamefont {Zhang},\ and\ \citenamefont
  {Wang}}]{chen2020tunable}%
  \BibitemOpen
  \bibfield  {author} {\bibinfo {author} {\bibfnamefont {G.}~\bibnamefont
  {Chen}}, \bibinfo {author} {\bibfnamefont {A.~L.}\ \bibnamefont {Sharpe}},
  \bibinfo {author} {\bibfnamefont {E.~J.}\ \bibnamefont {Fox}}, \bibinfo
  {author} {\bibfnamefont {Y.-H.}\ \bibnamefont {Zhang}}, \bibinfo {author}
  {\bibfnamefont {S.}~\bibnamefont {Wang}}, \bibinfo {author} {\bibfnamefont
  {L.}~\bibnamefont {Jiang}}, \bibinfo {author} {\bibfnamefont
  {B.}~\bibnamefont {Lyu}}, \bibinfo {author} {\bibfnamefont {H.}~\bibnamefont
  {Li}}, \bibinfo {author} {\bibfnamefont {K.}~\bibnamefont {Watanabe}},
  \bibinfo {author} {\bibfnamefont {T.}~\bibnamefont {Taniguchi}}, \bibinfo
  {author} {\bibfnamefont {Z.}~\bibnamefont {Shi}}, \bibinfo {author}
  {\bibfnamefont {T.}~\bibnamefont {Senthil}}, \bibinfo {author} {\bibfnamefont
  {D.}~\bibnamefont {Goldhaber-Gordon}}, \bibinfo {author} {\bibfnamefont
  {Y.}~\bibnamefont {Zhang}}, \ and\ \bibinfo {author} {\bibfnamefont
  {F.}~\bibnamefont {Wang}},\ }\href
  {https://www.nature.com/articles/s41586-020-2049-7} {\bibfield  {journal}
  {\bibinfo  {journal} {Nature}\ }\textbf {\bibinfo {volume} {579}},\ \bibinfo
  {pages} {56} (\bibinfo {year} {2020})}\BibitemShut {NoStop}%
\bibitem [{\citenamefont {Chen}\ \emph {et~al.}(2021)\citenamefont {Chen},
  \citenamefont {He}, \citenamefont {Zhang}, \citenamefont {Hsieh},
  \citenamefont {Fei}, \citenamefont {Watanabe}, \citenamefont {Taniguchi},
  \citenamefont {Cobden}, \citenamefont {Xu}, \citenamefont {Dean},\ and\
  \citenamefont {Yankowitz}}]{chen2020electrically}%
  \BibitemOpen
  \bibfield  {author} {\bibinfo {author} {\bibfnamefont {S.}~\bibnamefont
  {Chen}}, \bibinfo {author} {\bibfnamefont {M.}~\bibnamefont {He}}, \bibinfo
  {author} {\bibfnamefont {Y.-H.}\ \bibnamefont {Zhang}}, \bibinfo {author}
  {\bibfnamefont {V.}~\bibnamefont {Hsieh}}, \bibinfo {author} {\bibfnamefont
  {Z.}~\bibnamefont {Fei}}, \bibinfo {author} {\bibfnamefont {K.}~\bibnamefont
  {Watanabe}}, \bibinfo {author} {\bibfnamefont {T.}~\bibnamefont {Taniguchi}},
  \bibinfo {author} {\bibfnamefont {D.~H.}\ \bibnamefont {Cobden}}, \bibinfo
  {author} {\bibfnamefont {X.}~\bibnamefont {Xu}}, \bibinfo {author}
  {\bibfnamefont {C.~R.}\ \bibnamefont {Dean}}, \ and\ \bibinfo {author}
  {\bibfnamefont {M.}~\bibnamefont {Yankowitz}},\ }\href
  {https://doi.org/10.1038/s41567-020-01062-6} {\bibfield  {journal} {\bibinfo
  {journal} {Nature Physics}\ }\textbf {\bibinfo {volume} {17}},\ \bibinfo
  {pages} {374} (\bibinfo {year} {2021})}\BibitemShut {NoStop}%
\bibitem [{\citenamefont {Tsai}\ \emph {et~al.}(2019)\citenamefont {Tsai},
  \citenamefont {Zhang}, \citenamefont {Zhu}, \citenamefont {Luo},
  \citenamefont {Carr}, \citenamefont {Luskin}, \citenamefont {Kaxiras},\ and\
  \citenamefont {Wang}}]{tsai2019correlated}%
  \BibitemOpen
  \bibfield  {author} {\bibinfo {author} {\bibfnamefont {K.-T.}\ \bibnamefont
  {Tsai}}, \bibinfo {author} {\bibfnamefont {X.}~\bibnamefont {Zhang}},
  \bibinfo {author} {\bibfnamefont {Z.}~\bibnamefont {Zhu}}, \bibinfo {author}
  {\bibfnamefont {Y.}~\bibnamefont {Luo}}, \bibinfo {author} {\bibfnamefont
  {S.}~\bibnamefont {Carr}}, \bibinfo {author} {\bibfnamefont {M.}~\bibnamefont
  {Luskin}}, \bibinfo {author} {\bibfnamefont {E.}~\bibnamefont {Kaxiras}}, \
  and\ \bibinfo {author} {\bibfnamefont {K.}~\bibnamefont {Wang}},\ }\href
  {https://arxiv.org/abs/1912.03375} {\bibfield  {journal} {\bibinfo  {journal}
  {arXiv preprint arXiv:1912.03375}\ } (\bibinfo {year} {2019})}\BibitemShut
  {NoStop}%
\bibitem [{\citenamefont {Kariyado}\ and\ \citenamefont
  {Vishwanath}(2019)}]{kariyado2019flat}%
  \BibitemOpen
  \bibfield  {author} {\bibinfo {author} {\bibfnamefont {T.}~\bibnamefont
  {Kariyado}}\ and\ \bibinfo {author} {\bibfnamefont {A.}~\bibnamefont
  {Vishwanath}},\ }\href {https://doi.org/10.1103/PhysRevResearch.1.033076}
  {\bibfield  {journal} {\bibinfo  {journal} {Physical Review Research}\
  }\textbf {\bibinfo {volume} {1}},\ \bibinfo {pages} {033076} (\bibinfo {year}
  {2019})}\BibitemShut {NoStop}%
\bibitem [{\citenamefont {Hejazi}\ \emph {et~al.}(2020)\citenamefont {Hejazi},
  \citenamefont {Luo},\ and\ \citenamefont {Balents}}]{hejazi2020noncollinear}%
  \BibitemOpen
  \bibfield  {author} {\bibinfo {author} {\bibfnamefont {K.}~\bibnamefont
  {Hejazi}}, \bibinfo {author} {\bibfnamefont {Z.-X.}\ \bibnamefont {Luo}}, \
  and\ \bibinfo {author} {\bibfnamefont {L.}~\bibnamefont {Balents}},\ }\href
  {https://www.pnas.org/content/117/20/10721} {\bibfield  {journal} {\bibinfo
  {journal} {Proceedings of the National Academy of Sciences}\ }\textbf
  {\bibinfo {volume} {117}},\ \bibinfo {pages} {10721} (\bibinfo {year}
  {2020})}\BibitemShut {NoStop}%
\bibitem [{\citenamefont {Can}\ \emph {et~al.}(2021)\citenamefont {Can},
  \citenamefont {Tummuru}, \citenamefont {Day}, \citenamefont {Elfimov},
  \citenamefont {Damascelli},\ and\ \citenamefont {Franz}}]{can2020high}%
  \BibitemOpen
  \bibfield  {author} {\bibinfo {author} {\bibfnamefont {O.}~\bibnamefont
  {Can}}, \bibinfo {author} {\bibfnamefont {T.}~\bibnamefont {Tummuru}},
  \bibinfo {author} {\bibfnamefont {R.~P.}\ \bibnamefont {Day}}, \bibinfo
  {author} {\bibfnamefont {I.}~\bibnamefont {Elfimov}}, \bibinfo {author}
  {\bibfnamefont {A.}~\bibnamefont {Damascelli}}, \ and\ \bibinfo {author}
  {\bibfnamefont {M.}~\bibnamefont {Franz}},\ }\href {\doibase
  10.1038/s41567-020-01142-7} {\bibfield  {journal} {\bibinfo  {journal}
  {Nature Physics}\ }\textbf {\bibinfo {volume} {17}},\ \bibinfo {pages} {519}
  (\bibinfo {year} {2021})},\ \Eprint {http://arxiv.org/abs/2012.01412}
  {arXiv:2012.01412} \BibitemShut {NoStop}%
\bibitem [{\citenamefont {Volkov}\ \emph {et~al.}(2020)\citenamefont {Volkov},
  \citenamefont {Wilson},\ and\ \citenamefont {Pixley}}]{volkov2020magic}%
  \BibitemOpen
  \bibfield  {author} {\bibinfo {author} {\bibfnamefont {P.~A.}\ \bibnamefont
  {Volkov}}, \bibinfo {author} {\bibfnamefont {J.~H.}\ \bibnamefont {Wilson}},
  \ and\ \bibinfo {author} {\bibfnamefont {J.}~\bibnamefont {Pixley}},\ }\href
  {https://arxiv.org/abs/2012.07860} {\bibfield  {journal} {\bibinfo  {journal}
  {arXiv preprint arXiv:2012.07860}\ } (\bibinfo {year} {2020})}\BibitemShut
  {NoStop}%
\bibitem [{\citenamefont {May-Mann}\ and\ \citenamefont
  {Hughes}(2020)}]{may2020twisted}%
  \BibitemOpen
  \bibfield  {author} {\bibinfo {author} {\bibfnamefont {J.}~\bibnamefont
  {May-Mann}}\ and\ \bibinfo {author} {\bibfnamefont {T.~L.}\ \bibnamefont
  {Hughes}},\ }\href {https://doi.org/10.1103/PhysRevB.101.245126} {\bibfield
  {journal} {\bibinfo  {journal} {Physical Review B}\ }\textbf {\bibinfo
  {volume} {101}},\ \bibinfo {pages} {245126} (\bibinfo {year}
  {2020})}\BibitemShut {NoStop}%
\bibitem [{\citenamefont {Cano}\ \emph {et~al.}(2020)\citenamefont {Cano},
  \citenamefont {Fang}, \citenamefont {Pixley},\ and\ \citenamefont
  {Wilson}}]{cano2020moir}%
  \BibitemOpen
  \bibfield  {author} {\bibinfo {author} {\bibfnamefont {J.}~\bibnamefont
  {Cano}}, \bibinfo {author} {\bibfnamefont {S.}~\bibnamefont {Fang}}, \bibinfo
  {author} {\bibfnamefont {J.}~\bibnamefont {Pixley}}, \ and\ \bibinfo {author}
  {\bibfnamefont {J.~H.}\ \bibnamefont {Wilson}},\ }\href
  {https://arxiv.org/abs/2010.09726} {\bibfield  {journal} {\bibinfo  {journal}
  {arXiv preprint arXiv:2010.09726}\ } (\bibinfo {year} {2020})}\BibitemShut
  {NoStop}%
\bibitem [{\citenamefont {Wang}\ \emph
  {et~al.}(2020{\natexlab{b}})\citenamefont {Wang}, \citenamefont {Yuan},\ and\
  \citenamefont {Fu}}]{wang2020moir}%
  \BibitemOpen
  \bibfield  {author} {\bibinfo {author} {\bibfnamefont {T.}~\bibnamefont
  {Wang}}, \bibinfo {author} {\bibfnamefont {N.~F.~Q.}\ \bibnamefont {Yuan}}, \
  and\ \bibinfo {author} {\bibfnamefont {L.}~\bibnamefont {Fu}},\ }\href
  {https://arxiv.org/abs/2010.09753} {\bibfield  {journal} {\bibinfo  {journal}
  {arXiv preprint arXiv:2010.09753}\ } (\bibinfo {year}
  {2020}{\natexlab{b}})}\BibitemShut {NoStop}%
\bibitem [{\citenamefont {Gonz{\'a}lez-Tudela}\ and\ \citenamefont
  {Cirac}(2019)}]{gonzalez2019cold}%
  \BibitemOpen
  \bibfield  {author} {\bibinfo {author} {\bibfnamefont {A.}~\bibnamefont
  {Gonz{\'a}lez-Tudela}}\ and\ \bibinfo {author} {\bibfnamefont {J.~I.}\
  \bibnamefont {Cirac}},\ }\href {https://doi.org/10.1103/PhysRevA.100.053604}
  {\bibfield  {journal} {\bibinfo  {journal} {Physical Review A}\ }\textbf
  {\bibinfo {volume} {100}},\ \bibinfo {pages} {053604} (\bibinfo {year}
  {2019})}\BibitemShut {NoStop}%
\bibitem [{\citenamefont {Fu}\ \emph {et~al.}(2020)\citenamefont {Fu},
  \citenamefont {K{\"o}nig}, \citenamefont {Wilson}, \citenamefont {Chou},\
  and\ \citenamefont {Pixley}}]{fu2020magic}%
  \BibitemOpen
  \bibfield  {author} {\bibinfo {author} {\bibfnamefont {Y.}~\bibnamefont
  {Fu}}, \bibinfo {author} {\bibfnamefont {E.~J.}\ \bibnamefont {K{\"o}nig}},
  \bibinfo {author} {\bibfnamefont {J.~H.}\ \bibnamefont {Wilson}}, \bibinfo
  {author} {\bibfnamefont {Y.-Z.}\ \bibnamefont {Chou}}, \ and\ \bibinfo
  {author} {\bibfnamefont {J.~H.}\ \bibnamefont {Pixley}},\ }\href
  {https://doi.org/10.1038/s41535-020-00271-9} {\bibfield  {journal} {\bibinfo
  {journal} {npj Quantum Materials}\ }\textbf {\bibinfo {volume} {5}},\
  \bibinfo {pages} {1} (\bibinfo {year} {2020})}\BibitemShut {NoStop}%
\bibitem [{\citenamefont {Salamon}\ \emph {et~al.}(2020)\citenamefont
  {Salamon}, \citenamefont {Celi}, \citenamefont {Chhajlany}, \citenamefont
  {Fr{\'e}rot}, \citenamefont {Lewenstein}, \citenamefont {Tarruell},\ and\
  \citenamefont {Rakshit}}]{salamon2020simulating}%
  \BibitemOpen
  \bibfield  {author} {\bibinfo {author} {\bibfnamefont {T.}~\bibnamefont
  {Salamon}}, \bibinfo {author} {\bibfnamefont {A.}~\bibnamefont {Celi}},
  \bibinfo {author} {\bibfnamefont {R.~W.}\ \bibnamefont {Chhajlany}}, \bibinfo
  {author} {\bibfnamefont {I.}~\bibnamefont {Fr{\'e}rot}}, \bibinfo {author}
  {\bibfnamefont {M.}~\bibnamefont {Lewenstein}}, \bibinfo {author}
  {\bibfnamefont {L.}~\bibnamefont {Tarruell}}, \ and\ \bibinfo {author}
  {\bibfnamefont {D.}~\bibnamefont {Rakshit}},\ }\href
  {https://doi.org/10.1103/PhysRevLett.125.030504} {\bibfield  {journal}
  {\bibinfo  {journal} {Physical Review Letters}\ }\textbf {\bibinfo {volume}
  {125}},\ \bibinfo {pages} {030504} (\bibinfo {year} {2020})}\BibitemShut
  {NoStop}%
\bibitem [{\citenamefont {Luo}\ and\ \citenamefont
  {Zhang}(2021)}]{luo2021spin}%
  \BibitemOpen
  \bibfield  {author} {\bibinfo {author} {\bibfnamefont {X.-W.}\ \bibnamefont
  {Luo}}\ and\ \bibinfo {author} {\bibfnamefont {C.}~\bibnamefont {Zhang}},\
  }\href {https://doi.org/10.1103/PhysRevLett.126.103201} {\bibfield  {journal}
  {\bibinfo  {journal} {Physical Review Letters}\ }\textbf {\bibinfo {volume}
  {126}},\ \bibinfo {pages} {103201} (\bibinfo {year} {2021})}\BibitemShut
  {NoStop}%
\bibitem [{\citenamefont {Bistritzer}\ and\ \citenamefont
  {MacDonald}(2011{\natexlab{a}})}]{bistritzer2011moire}%
  \BibitemOpen
  \bibfield  {author} {\bibinfo {author} {\bibfnamefont {R.}~\bibnamefont
  {Bistritzer}}\ and\ \bibinfo {author} {\bibfnamefont {A.~H.}\ \bibnamefont
  {MacDonald}},\ }\href {https://doi.org/10.1073/pnas.1108174108} {\bibfield
  {journal} {\bibinfo  {journal} {Proceedings of the National Academy of
  Sciences}\ }\textbf {\bibinfo {volume} {108}},\ \bibinfo {pages} {12233}
  (\bibinfo {year} {2011}{\natexlab{a}})}\BibitemShut {NoStop}%
\bibitem [{\citenamefont {Bistritzer}\ and\ \citenamefont
  {MacDonald}(2011{\natexlab{b}})}]{bistritzer2011moire1}%
  \BibitemOpen
  \bibfield  {author} {\bibinfo {author} {\bibfnamefont {R.}~\bibnamefont
  {Bistritzer}}\ and\ \bibinfo {author} {\bibfnamefont {A.~H.}\ \bibnamefont
  {MacDonald}},\ }\href {https://doi.org/10.1103/PhysRevB.84.035440} {\bibfield
   {journal} {\bibinfo  {journal} {Physical Review B}\ }\textbf {\bibinfo
  {volume} {84}},\ \bibinfo {pages} {035440} (\bibinfo {year}
  {2011}{\natexlab{b}})}\BibitemShut {NoStop}%
\bibitem [{\citenamefont {Lopes~dos Santos}\ \emph {et~al.}(2012)\citenamefont
  {Lopes~dos Santos}, \citenamefont {Peres},\ and\ \citenamefont
  {Castro~Neto}}]{dos2012continuum}%
  \BibitemOpen
  \bibfield  {author} {\bibinfo {author} {\bibfnamefont {J.~M.~B.}\
  \bibnamefont {Lopes~dos Santos}}, \bibinfo {author} {\bibfnamefont
  {N.~M.~R.}\ \bibnamefont {Peres}}, \ and\ \bibinfo {author} {\bibfnamefont
  {A.~H.}\ \bibnamefont {Castro~Neto}},\ }\href
  {https://doi.org/10.1103/PhysRevB.86.155449} {\bibfield  {journal} {\bibinfo
  {journal} {Physical Review B}\ }\textbf {\bibinfo {volume} {86}},\ \bibinfo
  {pages} {155449} (\bibinfo {year} {2012})}\BibitemShut {NoStop}%
\bibitem [{\citenamefont {Shallcross}\ \emph {et~al.}(2010)\citenamefont
  {Shallcross}, \citenamefont {Sharma}, \citenamefont {Kandelaki},\ and\
  \citenamefont {Pankratov}}]{shallcross2010electronic}%
  \BibitemOpen
  \bibfield  {author} {\bibinfo {author} {\bibfnamefont {S.}~\bibnamefont
  {Shallcross}}, \bibinfo {author} {\bibfnamefont {S.}~\bibnamefont {Sharma}},
  \bibinfo {author} {\bibfnamefont {E.}~\bibnamefont {Kandelaki}}, \ and\
  \bibinfo {author} {\bibfnamefont {O.~A.}\ \bibnamefont {Pankratov}},\ }\href
  {https://doi.org/10.1103/PhysRevB.81.165105} {\bibfield  {journal} {\bibinfo
  {journal} {Physical Review B}\ }\textbf {\bibinfo {volume} {81}},\ \bibinfo
  {pages} {165105} (\bibinfo {year} {2010})}\BibitemShut {NoStop}%
\bibitem [{\citenamefont {Yin}\ \emph {et~al.}(2015)\citenamefont {Yin},
  \citenamefont {Qiao}, \citenamefont {Zuo}, \citenamefont {Li},\ and\
  \citenamefont {He}}]{yin2015experimental}%
  \BibitemOpen
  \bibfield  {author} {\bibinfo {author} {\bibfnamefont {L.-J.}\ \bibnamefont
  {Yin}}, \bibinfo {author} {\bibfnamefont {J.-B.}\ \bibnamefont {Qiao}},
  \bibinfo {author} {\bibfnamefont {W.-J.}\ \bibnamefont {Zuo}}, \bibinfo
  {author} {\bibfnamefont {W.-T.}\ \bibnamefont {Li}}, \ and\ \bibinfo {author}
  {\bibfnamefont {L.}~\bibnamefont {He}},\ }\href
  {https://doi.org/10.1103/PhysRevB.92.081406} {\bibfield  {journal} {\bibinfo
  {journal} {Physical Review B}\ }\textbf {\bibinfo {volume} {92}},\ \bibinfo
  {pages} {081406(R)} (\bibinfo {year} {2015})}\BibitemShut {NoStop}%
\bibitem [{\citenamefont {Anderson}(1987)}]{anderson1987resonating}%
  \BibitemOpen
  \bibfield  {author} {\bibinfo {author} {\bibfnamefont {P.~W.}\ \bibnamefont
  {Anderson}},\ }\href {https://doi.org/10.1126/science.235.4793.1196}
  {\bibfield  {journal} {\bibinfo  {journal} {science}\ }\textbf {\bibinfo
  {volume} {235}},\ \bibinfo {pages} {1196} (\bibinfo {year}
  {1987})}\BibitemShut {NoStop}%
\bibitem [{\citenamefont {Affleck}\ and\ \citenamefont
  {Marston}(1988)}]{affleck1988large}%
  \BibitemOpen
  \bibfield  {author} {\bibinfo {author} {\bibfnamefont {I.}~\bibnamefont
  {Affleck}}\ and\ \bibinfo {author} {\bibfnamefont {J.~B.}\ \bibnamefont
  {Marston}},\ }\href {https://doi.org/10.1103/PhysRevB.37.3774} {\bibfield
  {journal} {\bibinfo  {journal} {Physical Review B}\ }\textbf {\bibinfo
  {volume} {37}},\ \bibinfo {pages} {3774} (\bibinfo {year}
  {1988})}\BibitemShut {NoStop}%
\bibitem [{\citenamefont {Marston}\ and\ \citenamefont
  {Affleck}(1989)}]{marston1989large}%
  \BibitemOpen
  \bibfield  {author} {\bibinfo {author} {\bibfnamefont {J.~B.}\ \bibnamefont
  {Marston}}\ and\ \bibinfo {author} {\bibfnamefont {I.}~\bibnamefont
  {Affleck}},\ }\href {https://doi.org/10.1103/PhysRevB.39.11538} {\bibfield
  {journal} {\bibinfo  {journal} {Physical Review B}\ }\textbf {\bibinfo
  {volume} {39}},\ \bibinfo {pages} {11538} (\bibinfo {year}
  {1989})}\BibitemShut {NoStop}%
\bibitem [{\citenamefont {Wen}\ and\ \citenamefont
  {Lee}(1996)}]{wen1996theory}%
  \BibitemOpen
  \bibfield  {author} {\bibinfo {author} {\bibfnamefont {X.-G.}\ \bibnamefont
  {Wen}}\ and\ \bibinfo {author} {\bibfnamefont {P.~A.}\ \bibnamefont {Lee}},\
  }\href {https://doi.org/10.1103/PhysRevLett.76.503} {\bibfield  {journal}
  {\bibinfo  {journal} {Physical Review Letters}\ }\textbf {\bibinfo {volume}
  {76}},\ \bibinfo {pages} {503} (\bibinfo {year} {1996})}\BibitemShut
  {NoStop}%
\bibitem [{\citenamefont {Kim}\ and\ \citenamefont
  {Lee}(1999)}]{kim1999theory}%
  \BibitemOpen
  \bibfield  {author} {\bibinfo {author} {\bibfnamefont {D.~H.}\ \bibnamefont
  {Kim}}\ and\ \bibinfo {author} {\bibfnamefont {P.~A.}\ \bibnamefont {Lee}},\
  }\href {https://doi.org/10.1006/aphy.1998.5888} {\bibfield  {journal}
  {\bibinfo  {journal} {Annals of Physics}\ }\textbf {\bibinfo {volume}
  {272}},\ \bibinfo {pages} {130} (\bibinfo {year} {1999})}\BibitemShut
  {NoStop}%
\bibitem [{\citenamefont {Lee}\ \emph {et~al.}(2006)\citenamefont {Lee},
  \citenamefont {Nagaosa},\ and\ \citenamefont {Wen}}]{lee2006doping}%
  \BibitemOpen
  \bibfield  {author} {\bibinfo {author} {\bibfnamefont {P.~A.}\ \bibnamefont
  {Lee}}, \bibinfo {author} {\bibfnamefont {N.}~\bibnamefont {Nagaosa}}, \ and\
  \bibinfo {author} {\bibfnamefont {X.-G.}\ \bibnamefont {Wen}},\ }\href
  {https://doi.org/10.1103/RevModPhys.78.17} {\bibfield  {journal} {\bibinfo
  {journal} {Reviews of modern physics}\ }\textbf {\bibinfo {volume} {78}},\
  \bibinfo {pages} {17} (\bibinfo {year} {2006})}\BibitemShut {NoStop}%
\bibitem [{\citenamefont {Rantner}\ and\ \citenamefont
  {Wen}(2001)}]{rantner2001electron}%
  \BibitemOpen
  \bibfield  {author} {\bibinfo {author} {\bibfnamefont {W.}~\bibnamefont
  {Rantner}}\ and\ \bibinfo {author} {\bibfnamefont {X.-G.}\ \bibnamefont
  {Wen}},\ }\href {https://doi.org/10.1103/PhysRevLett.86.3871} {\bibfield
  {journal} {\bibinfo  {journal} {Physical review letters}\ }\textbf {\bibinfo
  {volume} {86}},\ \bibinfo {pages} {3871} (\bibinfo {year}
  {2001})}\BibitemShut {NoStop}%
\bibitem [{\citenamefont {Wen}(2002)}]{wen2002quantum}%
  \BibitemOpen
  \bibfield  {author} {\bibinfo {author} {\bibfnamefont {X.-G.}\ \bibnamefont
  {Wen}},\ }\href {https://doi.org/10.1103/PhysRevB.65.165113} {\bibfield
  {journal} {\bibinfo  {journal} {Physical Review B}\ }\textbf {\bibinfo
  {volume} {65}},\ \bibinfo {pages} {165113} (\bibinfo {year}
  {2002})}\BibitemShut {NoStop}%
\bibitem [{\citenamefont {Rantner}\ and\ \citenamefont
  {Wen}(2002)}]{rantner2002spin}%
  \BibitemOpen
  \bibfield  {author} {\bibinfo {author} {\bibfnamefont {W.}~\bibnamefont
  {Rantner}}\ and\ \bibinfo {author} {\bibfnamefont {X.-G.}\ \bibnamefont
  {Wen}},\ }\href {https://doi.org/10.1103/PhysRevB.66.144501} {\bibfield
  {journal} {\bibinfo  {journal} {Physical Review B}\ }\textbf {\bibinfo
  {volume} {66}},\ \bibinfo {pages} {144501} (\bibinfo {year}
  {2002})}\BibitemShut {NoStop}%
\bibitem [{\citenamefont {Hermele}\ \emph {et~al.}(2004)\citenamefont
  {Hermele}, \citenamefont {Senthil}, \citenamefont {Fisher}, \citenamefont
  {Lee}, \citenamefont {Nagaosa},\ and\ \citenamefont
  {Wen}}]{hermele2004stability}%
  \BibitemOpen
  \bibfield  {author} {\bibinfo {author} {\bibfnamefont {M.}~\bibnamefont
  {Hermele}}, \bibinfo {author} {\bibfnamefont {T.}~\bibnamefont {Senthil}},
  \bibinfo {author} {\bibfnamefont {M.~P.~A.}\ \bibnamefont {Fisher}}, \bibinfo
  {author} {\bibfnamefont {P.~A.}\ \bibnamefont {Lee}}, \bibinfo {author}
  {\bibfnamefont {N.}~\bibnamefont {Nagaosa}}, \ and\ \bibinfo {author}
  {\bibfnamefont {X.-G.}\ \bibnamefont {Wen}},\ }\href
  {https://doi.org/10.1103/PhysRevB.70.214437} {\bibfield  {journal} {\bibinfo
  {journal} {Physical Review B}\ }\textbf {\bibinfo {volume} {70}},\ \bibinfo
  {pages} {214437} (\bibinfo {year} {2004})}\BibitemShut {NoStop}%
\bibitem [{\citenamefont {Hermele}\ \emph {et~al.}(2005)\citenamefont
  {Hermele}, \citenamefont {Senthil},\ and\ \citenamefont {Fisher}}]{Hermele}%
  \BibitemOpen
  \bibfield  {author} {\bibinfo {author} {\bibfnamefont {M.}~\bibnamefont
  {Hermele}}, \bibinfo {author} {\bibfnamefont {T.}~\bibnamefont {Senthil}}, \
  and\ \bibinfo {author} {\bibfnamefont {M.~P.~A.}\ \bibnamefont {Fisher}},\
  }\href {https://doi.org/10.1103/PhysRevB.72.104404} {\bibfield  {journal}
  {\bibinfo  {journal} {Physical Review B}\ }\textbf {\bibinfo {volume} {72}},\
  \bibinfo {pages} {104404} (\bibinfo {year} {2005})}\BibitemShut {NoStop}%
\bibitem [{\citenamefont {Balents}(2019)}]{Balents}%
  \BibitemOpen
  \bibfield  {author} {\bibinfo {author} {\bibfnamefont {L.}~\bibnamefont
  {Balents}},\ }\href {https://scipost.org/SciPostPhys.7.4.048} {\bibfield
  {journal} {\bibinfo  {journal} {SciPost Phys}\ }\textbf {\bibinfo {volume}
  {7}},\ \bibinfo {pages} {48} (\bibinfo {year} {2019})}\BibitemShut {NoStop}%
\bibitem [{\citenamefont {Mora}\ \emph {et~al.}(2019)\citenamefont {Mora},
  \citenamefont {Regnault},\ and\ \citenamefont
  {Bernevig}}]{mora2019flatbands}%
  \BibitemOpen
  \bibfield  {author} {\bibinfo {author} {\bibfnamefont {C.}~\bibnamefont
  {Mora}}, \bibinfo {author} {\bibfnamefont {N.}~\bibnamefont {Regnault}}, \
  and\ \bibinfo {author} {\bibfnamefont {B.~A.}\ \bibnamefont {Bernevig}},\
  }\href {https://doi.org/10.1103/PhysRevLett.123.026402} {\bibfield  {journal}
  {\bibinfo  {journal} {Physical review letters}\ }\textbf {\bibinfo {volume}
  {123}},\ \bibinfo {pages} {026402} (\bibinfo {year} {2019})}\BibitemShut
  {NoStop}%
\bibitem [{\citenamefont {{Song}}\ \emph {et~al.}(2020)\citenamefont {{Song}},
  \citenamefont {{Lian}}, \citenamefont {{Regnault}},\ and\ \citenamefont
  {{Bernevig}}}]{Song2020TBGII}%
  \BibitemOpen
  \bibfield  {author} {\bibinfo {author} {\bibfnamefont {Z.-D.}\ \bibnamefont
  {{Song}}}, \bibinfo {author} {\bibfnamefont {B.}~\bibnamefont {{Lian}}},
  \bibinfo {author} {\bibfnamefont {N.}~\bibnamefont {{Regnault}}}, \ and\
  \bibinfo {author} {\bibfnamefont {B.~A.}\ \bibnamefont {{Bernevig}}},\
  }\href@noop {} {\bibfield  {journal} {\bibinfo  {journal} {arXiv e-prints}\
  ,\ \bibinfo {eid} {arXiv:2009.11872}} (\bibinfo {year} {2020})},\ \Eprint
  {http://arxiv.org/abs/2009.11872} {arXiv:2009.11872 [cond-mat.mes-hall]}
  \BibitemShut {NoStop}%
\bibitem [{Note1()}]{Note1}%
  \BibitemOpen
  \bibinfo {note} {As a side note, in the context of the algebraic spin liquid,
  the Dirac velocity anisotropy for the spinons, which is controlled by the
  deviation of $t/\Delta $ from 1, has been shown to be irrelevant in the
  renormalization group sense in the large-$N_f$ limit \cite
  {vafek2002relativity, franz2002qed, Hermele}. Here, $N_f$ refers to the
  number of flavor of spinons.}\BibitemShut {Stop}%
\bibitem [{\citenamefont {Guinea}\ \emph {et~al.}(2008)\citenamefont {Guinea},
  \citenamefont {Katsnelson},\ and\ \citenamefont
  {Vozmediano}}]{guinea2008midgap}%
  \BibitemOpen
  \bibfield  {author} {\bibinfo {author} {\bibfnamefont {F.}~\bibnamefont
  {Guinea}}, \bibinfo {author} {\bibfnamefont {M.~I.}\ \bibnamefont
  {Katsnelson}}, \ and\ \bibinfo {author} {\bibfnamefont {M.~A.~H.}\
  \bibnamefont {Vozmediano}},\ }\href
  {https://doi.org/10.1103/PhysRevB.77.075422} {\bibfield  {journal} {\bibinfo
  {journal} {Physical Review B}\ }\textbf {\bibinfo {volume} {77}},\ \bibinfo
  {pages} {075422} (\bibinfo {year} {2008})}\BibitemShut {NoStop}%
\bibitem [{\citenamefont {Wehling}\ \emph {et~al.}(2008)\citenamefont
  {Wehling}, \citenamefont {Balatsky}, \citenamefont {Tsvelik}, \citenamefont
  {Katsnelson},\ and\ \citenamefont {Lichtenstein}}]{wehling2008midgap}%
  \BibitemOpen
  \bibfield  {author} {\bibinfo {author} {\bibfnamefont {T.~O.}\ \bibnamefont
  {Wehling}}, \bibinfo {author} {\bibfnamefont {A.~V.}\ \bibnamefont
  {Balatsky}}, \bibinfo {author} {\bibfnamefont {A.~M.}\ \bibnamefont
  {Tsvelik}}, \bibinfo {author} {\bibfnamefont {M.~I.}\ \bibnamefont
  {Katsnelson}}, \ and\ \bibinfo {author} {\bibfnamefont {A.~I.}\ \bibnamefont
  {Lichtenstein}},\ }\href {https://doi.org/10.1209/0295-5075/84/17003}
  {\bibfield  {journal} {\bibinfo  {journal} {EPL (Europhysics Letters)}\
  }\textbf {\bibinfo {volume} {84}},\ \bibinfo {pages} {17003} (\bibinfo {year}
  {2008})}\BibitemShut {NoStop}%
\bibitem [{\citenamefont {Snyman}(2009)}]{snyman2009gapped}%
  \BibitemOpen
  \bibfield  {author} {\bibinfo {author} {\bibfnamefont {I.}~\bibnamefont
  {Snyman}},\ }\href {https://doi.org/10.1103/PhysRevB.80.054303} {\bibfield
  {journal} {\bibinfo  {journal} {Physical Review B}\ }\textbf {\bibinfo
  {volume} {80}},\ \bibinfo {pages} {054303} (\bibinfo {year}
  {2009})}\BibitemShut {NoStop}%
\bibitem [{\citenamefont {Tan}\ \emph {et~al.}(2010)\citenamefont {Tan},
  \citenamefont {Park},\ and\ \citenamefont {Louie}}]{tan2010graphene}%
  \BibitemOpen
  \bibfield  {author} {\bibinfo {author} {\bibfnamefont {L.~Z.}\ \bibnamefont
  {Tan}}, \bibinfo {author} {\bibfnamefont {C.-H.}\ \bibnamefont {Park}}, \
  and\ \bibinfo {author} {\bibfnamefont {S.~G.}\ \bibnamefont {Louie}},\ }\href
  {https://doi.org/10.1103/PhysRevB.81.195426} {\bibfield  {journal} {\bibinfo
  {journal} {Physical Review B}\ }\textbf {\bibinfo {volume} {81}},\ \bibinfo
  {pages} {195426} (\bibinfo {year} {2010})}\BibitemShut {NoStop}%
\bibitem [{\citenamefont {Tang}\ and\ \citenamefont {Fu}(2014)}]{Liang}%
  \BibitemOpen
  \bibfield  {author} {\bibinfo {author} {\bibfnamefont {E.}~\bibnamefont
  {Tang}}\ and\ \bibinfo {author} {\bibfnamefont {L.}~\bibnamefont {Fu}},\
  }\href {https://doi.org/10.1038/nphys3109} {\bibfield  {journal} {\bibinfo
  {journal} {Nature Physics}\ }\textbf {\bibinfo {volume} {10}},\ \bibinfo
  {pages} {964} (\bibinfo {year} {2014})}\BibitemShut {NoStop}%
\bibitem [{\citenamefont {Goerbig}(2009)}]{goerbig2009quantum}%
  \BibitemOpen
  \bibfield  {author} {\bibinfo {author} {\bibfnamefont {M.~O.}\ \bibnamefont
  {Goerbig}},\ }\href {https://arxiv.org/abs/0909.1998} {\bibfield  {journal}
  {\bibinfo  {journal} {arXiv preprint arXiv:0909.1998}\ } (\bibinfo {year}
  {2009})}\BibitemShut {NoStop}%
\bibitem [{\citenamefont {Vafek}\ \emph {et~al.}(2002)\citenamefont {Vafek},
  \citenamefont {Te{\v{s}}anovi{\'c}},\ and\ \citenamefont
  {Franz}}]{vafek2002relativity}%
  \BibitemOpen
  \bibfield  {author} {\bibinfo {author} {\bibfnamefont {O.}~\bibnamefont
  {Vafek}}, \bibinfo {author} {\bibfnamefont {Z.}~\bibnamefont
  {Te{\v{s}}anovi{\'c}}}, \ and\ \bibinfo {author} {\bibfnamefont
  {M.}~\bibnamefont {Franz}},\ }\href
  {https://doi.org/10.1103/PhysRevLett.89.157003} {\bibfield  {journal}
  {\bibinfo  {journal} {Physical review letters}\ }\textbf {\bibinfo {volume}
  {89}},\ \bibinfo {pages} {157003} (\bibinfo {year} {2002})}\BibitemShut
  {NoStop}%
\bibitem [{\citenamefont {Franz}\ \emph {et~al.}(2002)\citenamefont {Franz},
  \citenamefont {Te{\v{s}}anovi{\'c}},\ and\ \citenamefont
  {Vafek}}]{franz2002qed}%
  \BibitemOpen
  \bibfield  {author} {\bibinfo {author} {\bibfnamefont {M.}~\bibnamefont
  {Franz}}, \bibinfo {author} {\bibfnamefont {Z.}~\bibnamefont
  {Te{\v{s}}anovi{\'c}}}, \ and\ \bibinfo {author} {\bibfnamefont
  {O.}~\bibnamefont {Vafek}},\ }\href
  {https://doi.org/10.1103/PhysRevB.66.054535} {\bibfield  {journal} {\bibinfo
  {journal} {Physical Review B}\ }\textbf {\bibinfo {volume} {66}},\ \bibinfo
  {pages} {054535} (\bibinfo {year} {2002})}\BibitemShut {NoStop}%
\end{thebibliography}%

\appendix
\begin{widetext}
\section{General form of spin-independent interlayer tunneling}
\label{app:general_tunneling}
As we discussed in Sec. \ref{subsec:bootstrap}, the conditions in Eq. \eqref{eq:sym_Fourier} and Eq. \eqref{eq:exchange} relate the Fourier component $M_{\k}$ of the interlayer tunneling $M[\u]$ with other Fourier components within the set $\{M_{\k'} \, | \, \k' = \mathcal{M}_x^m \Rot^n(\k+\K) - \K ,m=0,1, n=0,1,2,3\}$ which contains either four or eight elements depending on the momentum $\k$. There is no constraint that relates the Fourier components of $M[\u]$ in different sets. Remember that the momentum $\k$ of the Fourier component $M_{\k}$ is restricted to $\k \in (\pi \mathbb{Z}, \pi \mathbb{Z})$ in the un-rotated coordinate as shown in Eq. \eqref{eq:Fourier}. 

When all momenta $\mathcal{M}_x^m \Rot^n(\k+\K) - \K$ for $m=0,1$ and $n=0,1,2,3$ are different, the set $\{M_{\k'} \, | \, \k' = \mathcal{M}_x^m \Rot^n(\k+\K) - \K ,m=0,1, n=0,1,2,3\}$ contains 8 different Fourier components of $M[\u]$. The most general spin-independent solution to the constraints in Eq. \eqref{eq:sym_Fourier} and Eq. \eqref{eq:exchange} is given by

\begin{align}
    M_{\k} & = w_{\k,1} \tau^1 + w_{\k,3} \tau^2 +  w_{\k,2} \tau^1 \mu^3 +  w_{\k,4}  \tau^2 \mu^3, \nonumber \\
    M_{\Rot(\k+\K) - \K} & = -w_{\k,3} \tau^1 + w_{\k,1} \tau^2 +  w_{\k,4} \tau^1 \mu^3 -  w_{\k,2}  \tau^2 \mu^3, \nonumber \\
    M_{\Rot^2(\k+\K) - \K} & = -w_{\k,1} \tau^1 - w_{\k,3} \tau^2 -  w_{\k,2} \tau^1 \mu^3 -  w_{\k,4}  \tau^2 \mu^3, \nonumber \\
    M_{\Rot^3(\k+\K) - \K} & = w_{\k,3} \tau^1 - w_{\k,1} \tau^2 -  w_{\k,4} \tau^1 \mu^3 +  w_{\k,2}  \tau^2 \mu^3, \nonumber \\
    M_{\mathcal{M}_x (\k+\K)-\K} & = w_{\k,3} \tau^1 + w_{\k,1} \tau^2 -  w_{\k,4} \tau^1 \mu^3 -  w_{\k,2}  \tau^2 \mu^3, \nonumber \\
    M_{\mathcal{M}_x \Rot(\k+\K) - \K} & =  w_{\k,1} \tau^1 - w_{\k,3}\tau^2 +  w_{\k,2} \tau^1 \mu^3 -  w_{\k,4}  \tau^2 \mu^3, \nonumber \\
     M_{\mathcal{M}_x \Rot^2(\k+\K) - \K} & = -w_{\k,3} \tau^1 - w_{\k,1} \tau^2 +  w_{\k,4} \tau^1 \mu^3 +  w_{\k,2}  \tau^2 \mu^3, \nonumber \\
     M_{\mathcal{M}_x \Rot^3(\k+\K) - \K} & = - w_{\k,1} \tau^1 + w_{\k,3} \tau^2 -  w_{\k,2} \tau^1 \mu^3 +  w_{\k,4}  \tau^2 \mu^3,
     \label{eq:M_GeneralSolution}
\end{align}
where $w_{\k,1}$, $w_{\k,2}$, $w_{\k,3}$, and $w_{\k,4}$ are real parameters that are not subject to further constraints.

The situation in which the set $\{M_{\k'} \, | \, \k' = \mathcal{M}_x^m \Rot^n(\k+\K) - \K ,m=0,1, n=0,1,2,3\}$ contains only 4 different Fourier components of $M[\u]$ occurs only when $ M_x (\k+\K) = \Rot (\k+\K) $ or $ M_x (\k+\K) = \Rot^3 (\k+\K) $. When $ M_x (\k+\K) = \Rot (\k+\K) $, the solution to the constraints in Eq. \eqref{eq:sym_Fourier} and Eq. \eqref{eq:exchange} is still given by Eq. \eqref{eq:M_GeneralSolution} but with an extra condition that
\begin{align}
    w_{\k,3} = w_{\k,4} = 0.
\end{align}
The minimal set of Fourier components of $M[\u]$ discussed in Sec. \ref{subsec:bootstrap}, which is the case with $\k=0$, exactly fits into this situation. When $ M_x (\k+\K) = \Rot^3 (\k+\K) $, the solution is given by Eq. \eqref{eq:M_GeneralSolution} but with an extra condition that
\begin{align}
    w_{\k,1} = w_{\k,2} = 0.
\end{align}

Notice that the general form of the spin-independent interlayer tunneling that satisfies the conditions Eq. \eqref{eq:sym_Fourier} and Eq. \eqref{eq:exchange} allows us to rewrite the Fourier expansion of $M[\u]$ as
\begin{align}
M[\u] & =\sum_{\k \in (\pi \mathbb{Z}, \pi \mathbb{Z} )} e^{i\k\cdot\u}M_{\k}  \nonumber \\
& = e^{-i\K \cdot \u } \sum_{\k \in (\pi \mathbb{Z}, \pi \mathbb{Z} )} i \sin\big( (\k+\K)\cdot \u \big) ~ M_{\k}
\label{eq:M_Fourier_Re}
\end{align}
This form of $M[\u]$ ensures that when $\u =0$, namely when the two layers of square lattices are not displaced relative to each other, the spin-independent interlayer tunneling vanishes. Also, we notice that the valley index $\mu^3 = \pm 1$ is a good quantum number under the most general spin-independent interlayer tunneling allowed by the conditions Eq. \eqref{eq:sym_Fourier} and Eq. \eqref{eq:exchange}. In the case where the two layers are rigidly twisted by a relative angle $\theta$, by knowing that $M_{\k}$'s are all Hermitian in Eq. \eqref{eq:M_Fourier_Re}, we can show straightforwardly that the quantum number $s$ introduced in Eq. \eqref{eq:ansatz} is also always a good quantum under the general form of $M[\u]$. When both $\mu^3$ and $s$ good quantum numbers, we can study the Hamiltonian of the twisted bilayer with both of them fixed like we did in Eq. \eqref{eq:period_B}. The general form of $M[\u]$ still preserves the $\RInv\Tt$ symmetry. Hence, the band structure for a fixed valley index $\mu^3$ and a fixed quantum number $s$ still has infinite connectivity.

\end{widetext}

\end{document}